\pdfoutput=1
\documentclass[a4paper, 11pt]{article}
\usepackage[utf8]{inputenc}
\usepackage[T1]{fontenc}
\usepackage{xspace}
\usepackage[english]{babel}
\usepackage{amsfonts, amsmath, amsthm, amssymb, mathrsfs, bbm, stmaryrd, mathtools}
\usepackage{graphicx}
\usepackage[left=2.5cm, right=2.5cm, top=2.5cm, bottom=2.5cm]{geometry}
\usepackage[colorlinks=true, citecolor=cyan, linkcolor=., urlcolor=.]{hyperref}
\usepackage{enumerate}
\usepackage[dvipsnames]{xcolor}
\usepackage{soul}
\usepackage{xfrac}
\usepackage{faktor}
\usepackage{bookmark}
\usepackage{authblk}

\usepackage{csquotes}
\usepackage[
	bibstyle=phys,
	biblabel=brackets,
	citestyle=numeric-comp,
	sorting=none,
	doi=false,
	eprint=true,
	pageranges=false,
	backend=biber
]{biblatex}

\DeclareFieldFormat[article,inproceedings,inbook]{title}{\mkbibitalic{#1\isdot}}
\AtEveryBibitem{
	\ifentrytype{book}{
		\clearfield{month}
	}{}
	\ifentrytype{online}{
		\clearfield{month}
	}{}
}

\DeclareBibliographyDriver{online}{%
	\usebibmacro{bibindex}%
	\usebibmacro{begentry}%
	\usebibmacro{author/editor+others/translator+others}%
	\setunit{\labelnamepunct}\newblock
	\printtext[title]{\printfield[titlecase]{title}}
	\printdate
	\printtext{,}
	\newunit\newblock
	\printfield[eprint:\strfield{eprinttype}]{eprint}
	\setunit{\bibpagerefpunct}\newblock
	\usebibmacro{pageref}
	\usebibmacro{finentry}}

\newcommand{\Tr}{\text{Tr}}

\newcommand{\ldv}[1]{\mathcal{L}_{#1}}
\newcommand{\dd}{\mathrm{d}}
\newcommand{\Sint}[1]{\int_\Sigma \dd \Sigma_{#1}\,}
\newcommand{\DSint}[2]{\int_{\partial\Sigma} \dd \sigma_{#1 #2}\,}

\newcommand{\ie}{i.e.\@\xspace}

\newcommand{\updown}[2]{^{#1}_{\phantom{#1}#2}}
\newcommand{\downup}[2]{_{#1}^{\phantom{#1}#2}}

\newcommand{\R}{\mathbbm{R}}

\newcommand{\D}{{\cal D}}
\newcommand{\G}{{\cal G}}
\newcommand{\V}{{\cal V}}
\newcommand{\W}{{\cal W}}
\newcommand{\T}{{\cal T}}
\newcommand{\If}{\mathfrak{i}}
\newcommand{\Is}{\iota}
\newcommand{\Lf}{\mathfrak{L}}

\newcommand{\defeq}{\vcentcolon=}
\newcommand{\X}{\mathtt{x}}
\newcommand{\Y}{\mathtt{y}}
\newcommand{\lem}{L_{\text{EM}}}

\newcommand\sbullet[1][.5]{\mathbin{\vcenter{\hbox{\scalebox{#1}{$\bullet$}}}}}

\numberwithin{equation}{section}

\begin{document}

\title{\LARGE{\bf{On the covariant formulation of gauge theories with boundaries}}}
\author[1]{Mehdi Assanioussi \thanks{mehdi.assanioussi@fuw.edu.pl}}
\author[2,1]{Jerzy Kowalski-Glikman \thanks{jerzy.kowalski-glikman@uwr.edu.pl}}
\author[1]{Ilkka M\"akinen \thanks{ilkka.makinen@ncbj.gov.pl}}
\author[1]{Ludovic Varrin \thanks{ludovic.varrin@ncbj.gov.pl}}
\affil[1]{National Centre for Nuclear Research, Pasteura 7, 02-093 Warsaw, Poland}
\affil[2]{Faculty of Physics and Astronomy, University of Wroclaw, Pl. Maksa Borna 9, 50-204 Wroclaw, Poland}
\date{}

\maketitle

\begin{abstract}
	In the present article, we review the classical covariant formulation of Yang--Mills theory and general relativity in the presence of spacetime boundaries, focusing mainly on the derivation of the presymplectic forms and their properties. We further revisit the introduction of the edge modes and the conditions which justify them, in the context where only field-independent gauge transformations are considered. We particularly show that the presence of edge modes is not justified by gauge invariance of the presymplectic form, but rather by the condition that the presymplectic form is degenerate on the initial field space, which allows to relate this presymplectic form to the symplectic form on the gauge reduced field space via pullback.
\end{abstract}

\tableofcontents

\newpage

\section{Introduction}

The understanding of gauge theories and their symmetries in the presence of spacetime boundaries is a physically rich problem, and highly non-trivial. The idea that boundaries could play an important role in the physics of gauge theories first emerged in the condensed matter community with the discovery of the \textit{quantum Hall effect} in 1980 \cite{vonKlitzing:1980pdk,Laughlin:1981jd}. When considering a system of electrons confined in a finite two-dimensional surface, it can be shown that there exists chiral excitations at the edge (boundary) which are responsible for the quantized conductance characteristic of the quantum Hall effect. It is now well understood that this effect can be described by a $(2+1)$-dimensional $U(1)$ Chern--Simons theory where, to assure gauge invariance of the action in the presence of a boundary, an additional topological term must be introduced. This additional term produces non-trivial boundary dynamics responsible for these chiral excitations. \cite{zhang1989,Frohlich:1991wb,FROHLICH1991,BALACHANDRAN_1992,wen1992}. These dynamical fields located at the spatial boundary of a system are called edge modes, edge states or boundary modes. Even though it was not fully understood at the time, this was the first hint that gauge theories behave in a particular way in the presence of spatial boundaries. Note that, in the quantum Hall effect, the gauge symmetry itself is responsible for the edge modes and this has directly observable physical consequences. This observation brings us back to an age-old debate in the physics community: in what sense, if at all, are gauge symmetries physical? It is indeed often believed that gauge symmetries are pure redundancy of the system and thus they do not carry any physical information. This is further supported by the standard method to understand a symmetry and its physical implications: Noether's theorem \cite{Noether_1971}. Noether's theorem states that, to each continuous symmetry of the theory, there exists a conserved charge. It turns out that the charge of gauge symmetries vanishes, validating the earlier conclusion that they do not convey any physical information. However, we have just outlined the direct physical impact of gauge symmetry in Chern--Simons theory, so how can we reconcile these two observations? The key lies in recognizing that the existence of boundaries can give rise to non-vanishing charges associated with gauge symmetries, and these charges will be localized exclusively at the system's edge. In the case of the quantum Hall effect, the $U(1)$ gauge group acquires boundary charges corresponding to the electric charge of the edge modes.

The mathematical framework that we will use to formalise these statements is known as the \textit{covariant phase space} formalism. As the name indicates, this formalism was developed with the goal of describing the Hamiltonian dynamics and symplectic geometry of field theories in a covariant way. As we will discuss in section \ref{sec:covariant_phase_space}, this formalism gives a recipe to define a symplectic structure of field theory and enables the covariant computation of Noether charges associated with symmetries. We then have a clear way to define physical symmetries as the ones yielding a non-vanishing charge. Using this definition, it can be shown that some gauge symmetries generate non-vanishing Noether charges in the presence of boundaries, and therefore become physical symmetries.

The covariant phase space formalism was elaborated in the late 70's \cite{GAWEDZKI1972307,Kijowski:1973gi,Kijowski:1976ze} and put into its modern form by Iyer, Lee, Wald and Zoupas \cite{L.W.1990,Wald_1993,Iyer_1994,Iyer_1995,Wald_2000}. While maintaining explicit covariance is inherently motivated in gravitational theories, its application to gauge theories brings about an additional layer of understanding \cite{C.W.1987,Crnkovic:1987tz,zuckerman1987,Margalef-Bentabol:2020teu,G:2021xvv,BarberoG:2021cei,Margalef-Bentabol:2022zso,G:2022ger}. Using this formalism, Witten and Crnkovi\'c showed in \cite{C.W.1987} that, in absence of boundaries, one can define a degenerate closed two-form on the space of solutions of the field equations, for both Yang--Mills theory and general relativity (seen as a gauge theory of the diffeomorphism group), which is gauge invariant. This indicated that the space of solutions for gauge theories is generally not a symplectic manifold, i.e.~one does not have a symplectic form which by definition should be a closed non-degenerate two-form on the space of solutions. In a gauge theory however, the space of interest is the physical field space which is given by the space of solutions \textit{modulo} gauge transformations, and this raises the question regarding the existence of the symplectic structure on this space. The authors of \cite{C.W.1987} argue that to guarantee the existence of the symplectic form on the physical field space, it suffices for the closed two-form on the space of solutions to be the pullback of the desired symplectic form on the physical field space. They then show that this condition is equivalent to the vanishing of the charge associated with gauge transformations. This analysis establishes the consistency of the symplectic structures of gauge theories obtained through the covariant phase space when no spacetime boundaries are present.

Similar research played a significant role in solidifying the interest of the fundamental physics community in exploring the geometrical aspects of gauge theories and their interplay with asymptotic physics. One of the best known results in this direction is the derivation of the BMS (Bondi–Metzner–Sachs) group, describing the asymptotic symmetries of spacetime at null infinity \cite{Bondi:1962px,Sachs1962,Newman:1966ub}. While it was surprising at first to find an infinite dimensional group of physical symmetries in gravity, this is now well understood within the covariant phase space formalism \cite{ASHTEKAR1991}. This new understanding brought about a revolution in the asymptotic symmetry research and new results emerged \cite{Hogan:1985nyb,Christodoulou:1991cr,Brown:1992br,Ashtekar:1996cd,Adamo:2009fq,Barnich:2013sxa,Compere:2016jwb,Flanagan:2015pxa,Madler:2016xju,Barnich:2017ubf,Dolan:2018hfo,Bunster:2018yjr,Prabhu:2019fsp,Bunster:2019mup,Blanchet:2020ngx,Prabhu:2021cgk,Blanchet:2023pce}. Eventually, larger symmetry groups at conformal infinity were found, generalizing the BMS algebra \cite{Campiglia:2014yka,Hollands:2016oma,Compere:2018ylh,Flanagan:2019vbl,Campiglia:2020qvc,Freidel:2021fxf}. In the larger context of gauge symmetries, asymptotic symmetries have also been connected to soft theorems and memory effects in the seminal work by Strominger et al. \cite{Strominger:2013jfa,Cachazo:2014fwa,He:2014cra,Strominger:2014pwa,He:2014laa,Kapec:2015ena,Strominger:2017zoo,Pate:2017vwa}. Their \textit{infrared triangle} postulates a "triality" between these three seemingly unrelated topics in the infrared. Asymptotic symmetries represent a broad and compelling area of research, and they are instrumental in advancing our understanding of contemporary theoretical physics.

In recent years, the quest to find the most general symmetry group for gravity naturally led to the \textit{corner proposal} \cite{Freidel:2015gpa,D.F.2016,Gomes:2016mwl,Geiller:2017whh,Geiller:2017xad,Speranza2018LocalPS,Setare:2018mii,Gomes:2018dxs,Gomes:2018shn,Gomes:2019rgg,Gomes:2019xto,Riello:2019tad,Riello:2020zbk,Freidel2020EdgeMO1,Freidel2020EdgeMO2,Freidel2020EdgeMO3,Donnelly:2020xgu,Gomes:2020miz,Ciambelli:2021vnn,Riello:2021lfl,Donnelly:2022kfs,Sheikh-Jabbari:2022mqi,Adami:2023wbe,ciambelli2023asymptotic,Ciambelli:2023bmn}. In these works, it is shown that there exists a universal corner algebra (UCA) that is postulated to be the fundamental ingredient of both classical and quantum gravity. The BMS algebra and its generalizations are contained in this universal corner algebra. It is essential to highlight that the UCA encompasses not just asymptotic symmetries but also finite-distance ones. The UCA charges have support on a boundary located at a finite distance, as is the case for the quantum Hall effect discussed above. They are believed to play an important role in gauge theories, holography, black hole physics, quantum gravity and much more \cite{Carlip_1995,Ba_niados_1999,Carlip_2005,Barnich:2007bf,Compere:2008us,Engle_2010,Donnay:2016ejv,Hawking:2016msc,Afshar:2017okz,Carlip:2017xne,Adami:2020amw,Adami:2020ugu,Adami:2021sko,Adami:2021nnf,Adami:2021kvx,Adami:2022ktn,Adami:2023fbm,Park:2018xtt,Park:2022nkb,Grumiller:2022qhx}. One of the most intriguing areas in which they are believed to play a role is entanglement entropy. In their attempt to define localized subsystems in gauge theories and gravity, Donnelly and Freidel discovered a finite-distance boundary algebra of charges \cite{D.F.2016}. They showed that, when boundaries are introduced, edge modes are necessary to preserve the gauge invariance of the symplectic form with respect to field-dependent gauge transformations. This is accomplished by manually adding an edge mode term to the symplectic potential. These new boundary degrees of freedom can be acted upon by the surface symmetry group, which leaves the bulk fields invariant. One can then define the phase space associated with the union of two spatial hypersurfaces sharing a boundary in the following way: first, the individual phase spaces associated to the two regions are both extended by the introduction of the edge modes. Then, the phase space associated to the union of the two regions is defined as the quotient of the Cartesian product of the individual extended phase spaces, by the action of the surface symmetry group. This is sometimes called a fusion product. Intuitively, the dynamics of the edge modes at the boundary precisely takes care of the gluing of the two regions. In the quantum version of this fusion product, it means that tracing over the Hilbert space associated to one region gives the entanglement entropy between the two subsystems.

In the case of gravity, the analysis was initially restricted to diffeomorphisms preserving the boundary. This is because the contraction of the symplectic form with a vector field generating a diffeomorphism which changes the location of the boundary surface cannot be written as an exact differential. There is an additional corner term called the \textit{flux}. The presence of the flux term expresses the fact that those charges are not conserved or that the system is dissipative. However, it was later recognized in the corner proposal community that the flux term could be eliminated by including the spacetime embedding maps as additional phase space field variables of the theory. This is known as the \textit{extended phase space} formalism \cite{Freidel:2021cjp,Ciambelli:2021nmv,klinger2023extended}. The embedding maps then play the role of an edge mode, encoding the effect of the fluxes and restoring the integrability of the charge\footnote{By integrability, we mean here that the contraction of the symplectic form with the vector generating the symmetry is an exact differential.}. This extended formalism enables a cohesive approach to symmetries and their charges, whether the former preserves the boundary or not.

While it is clear that boundary symmetries play an important role in fundamental physics, the introduction of the edge modes can be subject to confusion. As mentioned earlier, they were initially proposed in \cite{D.F.2016} as a necessity for the symplectic form to be gauge invariant in the presence of boundaries. This is however needed only when considering gauge transformation with field-dependent parameters. It should also be noted that the expression of gauge invariance in their formulation is given by the vanishing of the charge associated to gauge transformations. This is very reminiscent of Witten and Crnkovi\'c's condition for the presymplectic form on the field space to be the pullback of a closed symplectic form on the gauge reduced field space. Furthermore, there is also a second notion of edge modes provided by the embedding maps in the extended phase space formalism. There, the role of the edge mode is to eliminate the flux term so that only the charge remains. It is clear that while the first edge modes are needed for all field-dependent gauge symmetries, the second one is only needed for spacetime symmetries and not internal gauge symmetries. How are these different objects connected if at all? What happens with field-independent gauge symmetries? What about global spacetime symmetries in covariant gauge theories? This work is aimed at clarifying the role of the various edge modes in gauge theories with boundaries, as well as describing how they arise. The second source of confusion arises from the highly mathematical and abstract framework used in a substantial portion of the literature on this subject, making it challenging for non-specialists to engage with. However, such an advanced formalism is not essential for grasping the core concepts discussed here. Subsequently, the following content has been presented in the simplest mathematical language possible, requiring minimal prerequisites. This approach is intended to make the topic more accessible to a wider audience.

In this article, we start by presenting the covariant phase space formalism in a simple language in section \ref{sec:covariant_phase_space}. We briefly review Cartan calculus on spacetime in order to introduce its generalization to the field space, which serves as the primary tool of the formalism. 
We first apply the formalism to Yang--Mills theory in section \ref{sec:YangMillsTheory}, and we show that the presymplectic form is invariant with respect to field-independent gauge transformations, even in the presence of a boundary. The edge modes introduced in \cite{D.F.2016} are thus not needed for that purpose. However, we also show that the presymplectic form produces a gauge symmetry charge with support
on the boundary. Therefore, in order to reinstate the degeneracy condition of \cite{C.W.1987}, we introduce an additional group-valued field, the edge mode, and we demonstrate how this field contributes to cancel the emergent charges on the boundaries.  
We then study the global Poincaré symmetry in Yang--Mills theory, exemplified by Maxwell theory, and show how the notions of fluxes emerge within the familiar context of electrodynamics. 
In section \ref{sec:gravityeinsteinhilbert}, we move to analyzing the gravitational theory defined by the Einstein--Hilbert action, with field-independent diffeomorphisms as a symmetry group. 
We show that, as in the case of global symmetries in Yang--Mills theory, fluxes are again present when transformations which do not preserve the boundary are considered. The restriction to diffeomorphisms preserving the boundary provides a charge supported on the boundary, and it corresponds to the well-known Komar charge \cite{Komar:1958wp}. We then illustrate these results by an example where we calculate the boundary charges for the Kerr–Newman–de Sitter spacetime and its limiting cases.
The presence of fluxes serves as a motivation to extend the covariant phase space formalism to accommodate symmetries which do not preserve spacetime boundaries. This is accomplished in section \ref{sec:extendedphasespace} by considering that spacetime regions are defined through embedding maps of some abstract space, these embedding maps are then included in the theory as new phase space variables. 
In both cases of Yang--Mills theory and Einstein--Hilbert gravity, we demonstrate that the variation of the extended actions produces an additional term on the boundary which cancels the fluxes and restores the integrability of the charges.
Thus, in the extended formalism, the fluxes are absorbed by the presymplectic form. Finally, we show that, in the case of gravity, the additional edge mode contribution required for satisfying the degeneracy condition can be expressed in terms of the embedding map. This implies that the edge mode in gravity \textit{is} the embedding map.
We conclude the article with a summary and some comments in section \ref{sec:summary}.

\section{Covariant phase space formalism}\label{sec:covariant_phase_space}

The covariant phase space formalism is a geometrical framework used to compute the charges and their algebra of a given theory without explicitly choosing a time coordinate and breaking covariance. This formalism is a mathematical construction that hinges on Anderson's variational bicomplex \cite{Anderson1991}. In these notes, however, we will focus on presenting the covariant phase space in an accessible way that does not require advanced prior knowledge of differential geometry. For modern and comprehensive reviews we refer the reader to \cite{Margalef-Bentabol:2020teu,Harlow_2020}. The main idea is to promote the notion of Cartan calculus to the space of field configurations, which will be defined more precisely below. We therefore start by a brief refresher of this notion on spacetime and then move on to the generalization to field space. We end the section with the description of the Lagrangian formalism and Noether's theorem in this newly introduced language.

\subsection{Cartan calculus on spacetime}

Given a $4$-dimensional spacetime $M$, let us denote the space of $p$-forms by $\Lambda^{p}(M)$ and  the space of all differential forms on the manifold by $\Lambda^{\sbullet}(M)$. The Cartan calculus consists of a set of three \textit{differentiations}: the exterior derivative $\dd$, the interior product $\Is_v$ and the Lie derivative $\ldv{v}$, where $v \in \mathrm{T}M$ is a spacetime vector field. The exterior derivative is a derivation of degree 1, which means that it raises the rank of the form by one
\begin{equation}
	\dd : \Lambda^{p}(M) \longrightarrow \Lambda^{p+1}(M).
\end{equation}
It is the antisymmetrisation of the usual differentiation operation on tensors and is defined as follows. Let $v_i$ be spacetime vector fields and $\alpha$ a $p$-form. Then
\begin{align}
	\dd\alpha(v_1,\ldots,v_{p+1}) &= \sum_{i=1}^p (-1)^{i+1}\ v_i\left(\alpha(v_1,\ldots,v_{i-1},v_{i+1},\ldots, v_{p+1})\right)\nonumber\\
	&+ \sum_{i<j} (-1)^{i+j}\ \alpha([v_i,v_j],v_1\ldots,v_{i-1},v_{i+1},\ldots,v_{j-1},v_{j+1}, v_{p+1})
\end{align}
where $[v_i,v_j]$ denotes the commutator of two vector fields.

On the other hand, the interior product is defined as the contraction of a differential form with a vector field $v$. It is a differentiation map of degree $-1$, which means that it lowers the rank of the form
\begin{equation}
	\Is_v: \Lambda^{p}(M) \longrightarrow \Lambda^{p-1}(M). 
\end{equation}
The interior product is sometimes called a \textit{contraction}, and its action on a $p$-form $\alpha$ is given by
\begin{equation}
	(\Is_v \alpha)(v_1,v_2,...,v_{p-1}) = \alpha(v,v_1,v_2,...,v_{p-1}).
\end{equation}
Finally, the Lie derivative can be given by a combination of the two former differentiations, called \textit{Cartan's magic formula}
\begin{align}
	\ldv{v} \alpha = \dd(\iota_v \alpha) +  \iota_v(\dd \alpha)  
\end{align}

The name "differentiation" comes from the fact that these operators obey a graded Leibniz rule with respect to the exterior product:
\begin{align}
	\dd (\alpha \wedge \beta) &= (\dd \alpha) \wedge \beta + (-1)^p \alpha \wedge (\dd \beta),  \label{spacetimeexteriorcalculus1} \\
	\Is_v (\alpha \wedge \beta) &= (\Is_v \alpha) \wedge \beta + (-1)^p \alpha \wedge (\Is_v \beta), \label{spacetimeexteriorcalculus2} \\
	\ldv{v} (\alpha \wedge \beta) &= (\ldv{v} \alpha) \wedge \beta + \alpha \wedge (\ldv{v} \beta), \label{spacetimeexteriorcalculus3}
\end{align}
for all $\alpha \in \Lambda^{p}(M)$ and $\beta \in \Lambda^{q}(M)$. Note that by definition $\dd^2 = \Is_v^2 = 0$, and
\begin{align}
	\left[ \ldv{v},\ldv{w} \right] &= \ldv{\left[ v,w \right]}, \\
	\Is_v \Is_w + \Is_w \Is_v &= 0 .
\end{align}
for all $v,w \in \mathrm{T}M$. One can then show that the following relations between the three operators hold
\begin{align}
	\left[\ldv{v},\Is_w\right] &= \Is_{\left[v,w\right]}, \label{differentationsrelation1} \\
	\left[\ldv{v},\dd \right] &= 0. \label{differentationsrelation2}
\end{align}

\subsection{Cartan calculus on field space}\label{sec2.2}

We now move to generalize the definitions and results of the previous section to the space $\mathcal{F}$ of all possible field configurations $\{\varphi^a(x)\}$ on spacetime of a given theory, where $x$ represents a spacetime point and $a$ stands for all the indices (spacetime, algebra, etc.) of the field. From now on the space $\mathcal{F}$ will be referred to as the field space.
The field space $\mathcal{F}$ can be viewed as an infinite dimensional manifold where each point corresponds to a field configuration $\varphi^a$ on spacetime. In other words, it is a manifold with coordinates $\varphi^a$ which consists of the collection of all field components at all spacetime points (see for example \cite{KrigleMichor1997} for more details). 

Vector fields $\T$ on $\mathcal{F}$, i.e.~$\T \in T\mathcal{F}$, can be expressed in a basis generated by the functional derivatives with respect to the field configurations $\varphi^a$:
\begin{equation}\label{fieldspacevector}
	\T = \int \dd^4 x \, \T^a(x) \frac{\delta}{\delta \varphi^a(x)},
\end{equation}
In particular, one can define vectors on the field space $\V$ which are tangent to gauge symmetry orbits when these are present. These vectors are associated to the generators of the gauge transformations, generically denoted $\epsilon$, and we have
\begin{equation}\label{fieldspacetangentvector}
	\V = \int \dd^4 x \, \Delta_\epsilon \varphi^a(x) \frac{\delta}{\delta \varphi^a(x)},
\end{equation}
where $\Delta_\epsilon \varphi^a$ stands for the infinitesimal variation of the field $\varphi^a$ under a gauge transformation. For instance, in the case of diffeomorphism symmetry, the tangent vectors of gauge orbits on the field space $\V$ are generated by spacetime vectors $\epsilon$ via the Lie derivative of the field configuration, namely
\begin{equation}\label{fieldspaceSpacetimevector}
	\V = \int \dd^4 x \, \ldv{\epsilon}\varphi^a(x) \frac{\delta}{\delta \varphi^a(x)}.
\end{equation}
The tangent vector $\V$ tells how the field changes under infinitesimal translation in spacetime.

We now want to define differential forms on $\mathcal{F}$ in order to introduce the Cartan calculus. The starting point of the covariant phase space formalism is to interpret the variation of a field as the field space exterior derivative $\delta$ acting on the field $\varphi$, and consequently producing a field space one-form:
\begin{equation}
	\delta : \mathcal{F} \longrightarrow \Lambda^{(1)}(\mathcal{F}).    
\end{equation}
This interpretation connects nicely to the usual meaning of the variation in a way that will soon become clear. The defining relation of the one-form $\delta \varphi^a$ is given by its action on the basis vector fields on $\mathcal{F}$
\begin{equation}
	\delta\varphi^a\left(\frac{\delta}{\delta \varphi^b}\right) = \delta^a_b.
\end{equation}
It follows that the action of $\delta \varphi^a$ on the vector $\V$ in \eqref{fieldspacetangentvector} produces the vector components as in the usual spacetime setup
\begin{equation}
	\delta \varphi^a (\V) = \Delta_\epsilon \varphi^a.
\end{equation}

We can further introduce the interior product $\If_\V$ on the space of one-forms on $\mathcal{F}$
. It maps the basis\footnote{Note that it suffices to define its action on $\delta \varphi^a$, as these one-forms constitute a basis of the space of all one-forms.} $\delta \varphi^a$ as
\begin{equation}
	\begin{aligned}
		\If_\V: \Lambda^{(1)}(\mathcal{F}) &\longrightarrow \mathcal{F}\\
		\delta \varphi^a &\mapsto \If_\V \delta \varphi^a = \Delta_\epsilon \varphi^a.
		\label{fieldspaceinteriorproduct}
	\end{aligned}
\end{equation}
The above property makes it clear how our field space exterior derivative connects to the usual interpretation of the variation of a field $\If_\V \delta \varphi^a = \delta \varphi^a (\V)$.

The last ingredient needed for Cartan calculus is a field space Lie derivative that we will denote $\Lf_\V$ and introduce through Cartan's magic formula:
\begin{equation}
	\Lf_\V = \delta \If_\V + \If_\V \delta.    
	\label{cartan}
\end{equation}
Finally, it is straightforward to introduce an exterior product, and generalize $\delta$ and $\If_\V$ to $p$-forms in such a way that the field space analogue of relations \eqref{spacetimeexteriorcalculus1}--\eqref{spacetimeexteriorcalculus3} holds. We then have a field space Cartan calculus with the standard relations
\begin{align}
	\left[\Lf_\V,\If_\W \right] &= \If_{[\V,\W]},\\
	\left[\Lf_\V,\delta\right] &= 0,
\end{align}
where, $\V$ and $\W$ are field space vectors. Note that from the expression \eqref{fieldspaceinteriorproduct}, it follows that
\begin{equation}\label{fieldspaceLiederivative}
	\mathfrak{L}_\V \varphi^a (x)=	\iota_{\V} \, \delta \varphi^a(x) = \Delta_\epsilon \varphi^a(x),
\end{equation}
where in the first equality we used the fact that $\varphi^a(x)$ is a field space zero-form, i.e.~$\iota_\V \varphi^a(x)=0$.

We conclude this section by pointing out that the introduction of Cartan calculus on the field space implies that, in the description of a theory, we now have two types of forms, those on spacetime and those on the field space. As we will see, the dynamical objects defining a theory will in general be $(p,q)$-forms, meaning a $p$-form on spacetime and a $q$-form on field space. We denote the space of $(p,q)$-forms by $\Lambda^{(p,q)}(M,\mathcal{F})$.

\subsection{Lagrangian formalism and Noether's theorem}\label{sec:Lagrangeformalism}

In this section, we apply the formalism developed in the previous sections to analyze a Lagrangian system, and we formulate Noether's second theorem for local symmetries. For more details on the subject and the proof of the theorem, we refer the reader to \cite{L.W.1990,Wald_1993,Iyer_1994,Iyer_1995,Wald_2000,Harlow_2020,ciambelli2023asymptotic}.

In order to define the action of a theory, we integrate the Lagrangian density over the spacetime manifold $M$, which from here on it is assumed to be a globally hyperbolic spacetime. The Lagrangian density is a spacetime $4$-form, also called a top form. Moreover, it is a functional of the fields and their derivatives, and is therefore a function on field space i.e.~$0$-form \footnote{The dependence of the Lagrangian density on derivatives of the field does not change this fact. A proper treatment of the first and higher order derivatives of the fields in a geometrical language requires the notion of \textit{jet bundles}. This goes beyond the purpose of these notes. For a definition and application of the jet bundles in the covariant phase space formalism, we refer the reader to \cite{Anderson1991,KrigleMichor1997}}. The action of the theory is thus defined as
\begin{equation}
	S = \int_M L \left[\varphi^a\right],
\end{equation}
where $L \in \Lambda^{(4,0)}(M,\mathcal{F})$. By using the usual Leibniz rule, the variation of the Lagrangian density can always be written in the following form
\begin{equation}\label{VariationLagrangian}
	\delta L[\varphi^a] = E_a \delta \varphi^a + \dd \theta[\varphi^a,\delta \varphi^a],
\end{equation}
where $E_a$ are the equations of motion and $\theta$ is a $(3,1)$-form, that is $\theta \in \Lambda^{(3,1)}(M,\mathcal{F})$, called the \textit{symplectic potential current}. The form $\theta$ contains all of the necessary information to provide a symplectic structure \cite{Arnold, SympGeo} for the field space of the theory. By integrating $\theta$ on a submanifold $\Sigma$ of a Cauchy surface we obtain
\begin{equation}\label{SympPot}
	\Theta = \int_\Sigma \theta,
\end{equation}
which is a $(0,1)$-form called the \textit{symplectic potential (or tautological form)} \cite{SympGeo}. We can now take its field space exterior derivative to get a closed $(0,2)$-form
\begin{equation}\label{PresympForm}
	\Omega = \delta \Theta = \int_\Sigma \delta \theta,
\end{equation}
called the \textit{presymplectic form}.
It is important to mention that, under specific conditions, the presymplectic form $\Omega$ is independent of the choice of the submanifold $\Sigma$. This can be established by noticing that when the equations of motion are satisfied, i.e.~on-shell, $\delta \theta$ is a closed spacetime form: $\dd\delta\theta = 0$. It follows that given a compact spacetime region $M$ with boundary $\partial M$ we have
\begin{align}
	\int_M \dd\delta \theta = \int_{\partial M} \delta \theta = 0 .
\end{align}
By decomposing the boundary as $\partial M = \Sigma_1 \cup \Sigma_2 \cup \Gamma$, where $\Sigma_1$ and $\Sigma_2$ are future oriented compact submanifolds of two distinct Cauchy surfaces, with $\Sigma_2$ in the future of $\Sigma_1$, and $\Gamma$ a timelike open submanifold, we obtain
\begin{align}
	\int_{\partial M} \delta \theta = \int_{\Sigma_2} \delta \theta - \int_{\Sigma_1} \delta \theta + \int_{\Gamma} \delta \theta = 0 .
\end{align}
The presymplectic form $\Omega$ would be independent of the choice of submanifold if we manage to prove that
\begin{align}
	\int_{\Sigma_1} \delta \theta = \int_{\Sigma_2} \delta \theta ,
\end{align}
since the choice of spacetime region is arbitrary. In order to prove this result, we must have 
\begin{align}
	\int_{\Gamma} \delta \theta = 0 .
\end{align}
Because $\Gamma$ does not have a boundary, the most general condition on $\theta$ which satisfies this requirement is
\begin{align}\label{dC}
	\delta \theta|_{\Gamma} = \dd C ,
\end{align}
for some $C \in \Lambda^{(2,2)}(M,\mathcal{F})$, see \cite{Harlow_2020} for a discussion. Equation \eqref{dC} represents a sufficient condition to guarantee that the presymplectic form $\Omega$ is independent of the choice of compact submanifold or Cauchy surface. It is to be understood as a condition on the field configurations satisfying the equations of motion. In this article, we always assume that the condition \eqref{dC} is satisfied on-shell. 

Note that we use the term "presymplectic", instead of "symplectic", to indicate that the form $\Omega$ could be degenerate. For instance, this is often the case in gauge theories without boundaries: the presymplectic form is degenerate along the gauge symmetry directions, which is a result that can be understood as a consequence of Noether's second theorem \cite{Noether_1971}, as we will see later. This fact means that one cannot invert the presymplectic form to define a Poisson bracket on $\mathcal{F}$. However, this does not mean that one is unable to promote $\mathcal{F}$ to a phase space, because one can still introduce the Poisson bracket on the field space in the standard way. Namely, by first identifying the variables and their conjugate momenta, then imposing the canonical Poisson commutation rule. This approach leads to a definition of the phase space as a Poisson manifold \cite{Arnold, Vaisman}, but not necessarily a symplectic one. The points of degeneracy of the presymplectic form would then correspond to singular points of the Poisson bracket. Conceptually, the most important structure in the phase space analysis is the gauge reduced field space
\begin{equation}
	\tilde{\mathcal{F}} \defeq \mathcal{F}/G,
\end{equation}
where $G$ is the gauge group. The space $\tilde{\mathcal{F}}$ can be endowed with a symplectic (non-degenerate) form $\tilde{\Omega}$ in such a way that $\Omega$ is the pullback of $\tilde\Omega$ from $\tilde{\mathcal{F}}$ to $\mathcal{F}$. The symplectic form $\tilde{\Omega}$ would then induce a Poisson bracket on $\tilde{\mathcal{F}}$ which would coincide with the Poisson bracket inherited from the Poisson manifold introduced in the standard fashion. In fact, the degeneracy of the presymplectic form $\Omega$ on $\mathcal{F}$ is a necessary condition to make the gauge reduced field space $\tilde{\Omega}$ a symplectic manifold whose symplectic form provides the form $\Omega$ on $\mathcal{F}$ via pullback (see \cite{C.W.1987}). Ensuring this relation between the symplectic structures on $\tilde{\mathcal{F}}$ and $\mathcal{F}$ is central in our approach to give rise to the so-called edge modes in the context where spacetime boundaries are present. We will indeed show that imposing the degeneracy of the presymplectic form on $\mathcal{F}$ along the gauge directions, when a spacetime boundary is present, requires the introduction of new fields associated to the boundary. These fields are what we call edges modes, and they coincide with the ones introduced earlier in the literature, e.g.~\cite{D.F.2016}. 

The field space forms $\theta$, $\Theta$ and $\Omega$ encode the gauge invariant quantities associated to the symmetries of a given theory, these are the symmetry charges. In covariant gauge theories, one can distinguish two categories of symmetries: internal gauge symmetries, and spacetime symmetries (including field independent diffeomorphisms and possible global spacetime symmetries). 
As it turns out, the aforementioned field space forms are all invariant under internal (field independent) gauge transformations. In the case of spacetime symmetries, and in the absence of spacetime boundaries or in the presence of boundaries (even a boundary at infinity with boundary conditions) which are preserved by the symmetry transformation under consideration, only the symplectic potential $\Theta$ and the presymplectic form $\Omega$ are invariant. However, these forms fail to be invariant under the action of a generic spacetime transformation.

The three field space forms allow a natural derivation of the Noether current and the charges associated to the considered gauge transformation.
To illustrate this, consider a vector field $\V$ tangent to a gauge orbit. 
Using Cartan's magic formula on the symplectic potential in \eqref{SympPot}, we get
\begin{equation}
	\mathfrak{L}_\V \Theta = \If_\V \delta \Theta + \delta \If_\V \Theta = \If_\V \Omega + \int_\Sigma\delta \If_\V \theta. 
\end{equation}
where we used \eqref{PresympForm}. This implies
\begin{equation}\label{IOmega}
	\If_\V \Omega = \mathfrak{L}_\V \Theta -\delta \int_\Sigma \If_\V \theta. 
\end{equation}
As mentioned above, when one deals with an internal gauge symmetry, one has
\begin{equation}
	\mathfrak{L}_\V \Theta = 0, 
\end{equation}
and in this case one can write
\begin{equation}\label{hamiltonequation}
	\If_\V \Omega = \delta H_g[\V].
\end{equation}
where $H_g[\V]$ is the symmetry charge associated to the gauge transformation generated by the vector field $\V$ defined as
\begin{equation}\label{chargedefinition}
	H_g[\V] := \int_\Sigma J_\V.
\end{equation}
with $J_{\V}$ being the Noether current associated to the gauge transformation and given by
\begin{equation} \label{noethercurrentgauge}
	J_{\V} := - \If_\V \theta.
\end{equation}

In the case of a spacetime transformation, the term $\mathfrak{L}_\V \Theta$ does not necessarily vanish. Suppose that the field space vector $\V$ is associated to the spacetime vector field $v$ as in \eqref{fieldspaceSpacetimevector}. Using equation \eqref{fieldspaceLiederivative}, we can see that
\begin{equation}
	\mathfrak{L}_\V \Theta = \ldv{v} \Theta = \int_\Sigma \ldv{v} \theta = \int_\Sigma (\dd \Is_v \theta + \Is_v \dd \theta). 
\end{equation}
Then thanks to equation \eqref{VariationLagrangian}, and assuming that the equations of motion are satisfied, i.e.~we work on-shell, we can write
\begin{equation}
	\Is_v \dd \theta = \delta \Is_v L. 
\end{equation}
and therefore we have
\begin{equation}\label{IOmega2}
	\If_\V \Omega = \delta \int_\Sigma (\Is_v L - \If_\V \theta) + \int_\Sigma \dd \Is_v \theta. 
\end{equation}
In the presence of spacetime boundaries, the term $\int_\Sigma \dd \Is_v \theta$ induces the so-called fluxes through the boundary of $\Sigma$, and we have:
\begin{equation}
	\int_\Sigma \dd \Is_v \theta = \int_{\partial\Sigma} \Is_v \theta .
\end{equation}
But otherwise $\Sigma$ has no boundary and this term vanishes. Either way, the Noether current $J_{\V}$ associated to the spacetime symmetry transformation generated by $\V$ can be defined as
\begin{equation} \label{noethercurrentspacetime}
	J_{\V} := \Is_v L - \If_\V \theta.
\end{equation}
and the corresponding spacetime symmetry charge $H_s[\V]$ is then
\begin{equation}
	H_s[\V] := \int_\Sigma J_{\V} = \int_\Sigma (\Is_v L - \If_\V \theta). 
\end{equation}

Now that the charges are defined, we would want to introduce a Poisson bracket for these charges and obtain the charges algebra. However, before doing so, let us go back to the question of degeneracy of the presymplectic form. 
Noether's second theorem \cite{Noether_1971} states that, for local symmetries, the current $J_{\V}$ is in general an exact differential on spacetime, namely
\begin{equation}\label{noether2theorem}
	J_\V = \dd Q_\V.
\end{equation}
This implies that when no boundaries are present, the associated charge vanishes 
\begin{equation}
	H[\V] = \int_\Sigma \dd Q_\V = 0, 
\end{equation}
and we get
\begin{equation} \label{degeneracy}
	\If_\V \Omega = 0.
\end{equation}
This is the situation that we discussed earlier in this section in the context where no boundaries are present: the presymplectic form is degenerate in the gauge directions.
In the presence of boundaries, the situation differs. Firstly, the introduction of a boundary implies through equations \eqref{chargedefinition} and \eqref{noether2theorem} that the gauge symmetry has now a charge with support on the boundary $\partial \Sigma$ of $\Sigma$:
\begin{equation}
	H[\V] = \int_{\partial \Sigma} Q_\V.
\end{equation}
Consequently this may give a non vanishing contribution to the right hand side of equation \eqref{IOmega}. As mentioned earlier, ensuring that we have a symplectic manifold as a reduced phase space requires a degenerate presymplectic form on $\mathcal{F}$, and consequently the addition of new degrees of freedom corresponding to the edge modes.

Note however that in general we can still introduce a Poisson bracket to define the algebra of the symmetry charges, provided that the term $\dd \Is_v \theta$ in \eqref{IOmega2} which induces the fluxes through the boundary vanishes. These brackets can be defined as
\begin{equation}\label{ChargesAlg}
	\{H[\V],H[\W]\} := \mathfrak{L}_\V H[\W] = \If_{\V} \If_{\W} \Omega ,
\end{equation}
where $\V$ and $\W$ are arbitrary vector fields on the field space. We will see later, and as shown in \cite{Ciambelli:2021vnn, Ciambelli:2021nmv}, that this condition of vanishing fluxes can be realized in generally covariant gauge theories by considering an extension of the phase space of the theory with boundaries. Namely, the spacetime embedding maps are promoted to be additional field space variables.

This concludes this section and we now move to applying the framework introduced above to the cases of Yang--Mills on Minkowski spacetime and Einstein metric gravity.

\section{Yang--Mills theory}\label{sec:YangMillsTheory}

Yang--Mills theory on Minkowski spacetime with a gauge group $G$ is defined by the action
\begin{equation}
	S_\text{YM} [A]:= -\frac{1}{4}\int_M \dd^4x\,\Tr\bigl(F^{\mu\nu}F_{\mu\nu}\bigr) ,
	\label{YMaction}
\end{equation}
where
\begin{equation}
	F_{\mu\nu} := \partial_\mu A_\nu - \partial_\nu A_\mu + [A_\mu, A_\nu]
	\label{F}
\end{equation}
is the field tensor, or the curvature of the connection $A_\mu$, both valued in the Lie algebra  $\G$ of $G$. 

Consider the variation of the action,
\begin{equation}
	\delta S_\text{YM} [A, \delta A]= -\frac{1}{2}\int_M \dd^4x\,\Tr\bigl(F^{\mu\nu}\delta F_{\mu\nu}\bigr) .
\end{equation}
Using the identity
\begin{equation}
	\delta F_{\mu\nu} = \D_\mu\delta A_\nu - \D_\nu\delta A_\mu ,
\end{equation}
where $\D_\mu\alpha := \partial_\mu\alpha + [A_\mu, \alpha]$ denotes the gauge covariant derivative for any $\G$-valued quantity $\alpha$, we find
\begin{equation}
	\delta S_\text{YM} = \int_M \dd^4x\,\Tr\Bigl(\bigl(\D_\mu F^{\mu\nu}\bigr)\delta A_\nu\Bigr) - \int_M \dd^4x\,\partial_\mu\Bigl(\Tr\bigl(F^{\mu\nu}\delta A_\nu\bigr)\Bigr).
	\label{dS_YM}
\end{equation}
The first term in \eqref{dS_YM} represents Yang--Mills equations of motion
\begin{equation}
	\D_\mu F^{\mu\nu} = 0.
	\label{EOM_YM}
\end{equation}
The second term provides the symplectic potential
\begin{equation}
	\Theta_\text{YM} [A, \delta A] := \int_\Sigma \dd\Sigma_\mu\,\theta_\text{YM}^\mu ,
	\label{ThetaYM}
\end{equation}
where $\theta_\text{YM}^\mu$ is the symplectic potential current given by
\begin{equation}
	\theta_\text{YM}^\mu [A, \delta A] := -\Tr\bigl(F^{\mu\nu}\delta A_\nu\bigr) ,
	\label{thetaYM}
\end{equation}
and the integral is taken over a submanifold $\Sigma$ of a Cauchy surface in $M$, with $\dd\Sigma_\mu := \tfrac{1}{3!}\varepsilon_{\mu\nu\rho\sigma}\, \dd x^\nu \wedge \dd x^\rho \wedge \dd x^\sigma$ being the natural volume form on $\Sigma$ and $\varepsilon$ being the Levi-Civita symbol.

\subsection{Presymplectic form and its properties}

The presymplectic form $\Omega_\text{YM}$ is obtained as the exterior derivative of the symplectic potential on the field space with the operator $\delta$:
\begin{equation}
	\Omega_\text{YM} [A, \delta A] := \delta\Theta_\text{YM} [A, \delta A] = \int_\Sigma \dd\Sigma_\mu\,\omega_\text{YM}^\mu
	\label{OmegaYM}
\end{equation}
with $\omega_\text{YM}^\mu$ being the \textit{symplectic current} given by
\begin{equation}
	\omega_\text{YM}^\mu [A, \delta A] := \delta\theta_\text{YM}^\mu [A, \delta A] = -\Tr\bigl(\delta F^{\mu\nu}\delta A_\nu\bigr).
	\label{omegaYM}
\end{equation}

Under a local gauge transformation, described by a gauge function $g(x)$ valued in the gauge group $G$, the connection and the field tensor transform as
\begin{align}
	A_\mu &\to gA_\mu g^{-1} + g\partial_\mu g^{-1} \label{g*A} , \\
	F_{\mu\nu} &\to gF_{\mu\nu}g^{-1} \label{g*F} .
\end{align}
Applying the operator $\delta$ to these equations, and assuming that the gauge function $g$ is independent of the fields ($\delta g = 0$), we find that the field space one-forms $\delta A_\mu$ and $\delta F_{\mu\nu}$ transform homogeneously, \ie
\begin{align}
	\delta A_\mu &\to g\delta A_\mu g^{-1} \label{g*DA} , \\
	\delta F_{\mu\nu} &\to g\delta F_{\mu\nu}g^{-1} \label{g*DF} .
\end{align}
Inserting \eqref{g*F} and \eqref{g*DA} into \eqref{thetaYM}, we immediately see that the symplectic potential current $\theta_\text{YM}^\mu$ is gauge invariant:
\begin{equation}
	\theta_\text{YM}^\mu[A_g, \delta A_g] = \theta_\text{YM}^\mu[A, \delta A] ,
\end{equation}
where $A_g$ denotes the gauge transformed connection,
\begin{equation}
	A_g := gAg^{-1} + g(dg^{-1}).
\end{equation}
Similarly, using \eqref{g*DA} and \eqref{g*DF} in \eqref{omegaYM}, we establish the gauge invariance of the symplectic current $\omega^\mu$:
\begin{equation}
	\omega_\text{YM}^\mu[A_g, \delta A_g] = \omega_\text{YM}^\mu[A, \delta A].
\end{equation}
It follows from the above that the symplectic potential $\Theta_\text{YM}$ and the presymplectic form $\Omega_\text{YM}$ are gauge invariant, assuming field-independent gauge transformations, regardless of the presence or absence of any boundary of $\Sigma$.

\subsection{Yang--Mills charges and their algebra}

In the context of a field theory on a spacetime without boundaries, the charges generated by the presymplectic form \eqref{OmegaYM} vanish. However, in the presence of a spacetime boundary, we expect the presymplectic form \eqref{OmegaYM} to generate non vanishing boundary charges associated to the gauge symmetry of the theory, as discussed in section \ref{sec:Lagrangeformalism}. In order to calculate these charges for Yang--Mills theory, we need to evaluate the interior product of $\Omega_\text{YM}$ with an arbitrary vector field $\V$ tangent to a gauge orbit in the field space, and we proceed as follows.

Consider a functional ${\cal O}$ on the field space. The infinitesimal variation of ${\cal O}$ under an infinitesimal gauge transformation generated by a vector field $\V$ tangent to a gauge orbit in the field space is given by the Lie derivative of ${\cal O}$, namely $\Lf_\V{\cal O}$. In the case where the functional ${\cal O}$ is a $0$-form on the field space, and as a consequence of the Cartan's formula \eqref{cartan}, the Lie derivative of ${\cal O}$ reduces to the interior product of $\delta {\cal O}$ with the vector field $\V$:
\begin{align}
	\Lf_\V{\cal O} = \If_\V \delta {\cal O} ,
\end{align}  
because in this case $\If_\V {\cal O} = 0$. As discussed in section \ref{sec2.2}, it follows that the contraction of the fundamental one-forms $\delta A_\mu$ and $\delta F_{\mu\nu}$ with $\V$ gives the infinitesimal gauge variation of the fields $A_\mu$ and $F_{\mu\nu}$, namely
\begin{align}
	\If_\V\delta A_\mu &= \D_\mu\epsilon \label{iDA} , \\
	\If_\V\delta F_{\mu\nu} &= [F_{\mu\nu}, \epsilon] , \label{iDF}
\end{align}
where $\epsilon \in {\G}$ is the generator of the gauge transformation.
Using \eqref{iDA} and \eqref{iDF}, we may compute the contraction of the presymplectic form as
\begin{align}
	\If_\V\Omega_\text{YM} &= -\int_\Sigma \dd\Sigma_\mu\,\Tr\Bigl(\bigl(\If_\V\delta F^{\mu\nu}\bigr)\delta A_\nu - \delta F^{\mu\nu}\bigl(\If_\V\delta A_\nu\bigr)\Bigr) \notag \\
	&= -\int_\Sigma \dd\Sigma_\mu\,\Tr\Bigl([F^{\mu\nu}, \epsilon]\delta A_\nu - \delta F^{\mu\nu}\D_\nu\epsilon\Bigr) \notag \\
	&= \int_\Sigma \dd\Sigma_\mu\,\Tr\Bigl(\D_\nu\bigl(\delta F^{\mu\nu}\epsilon\bigr) - \bigl(\D_\nu\delta F^{\mu\nu}\bigr)\epsilon - \bigl[\delta A_\nu, F^{\mu\nu}\bigr] \epsilon\Bigr) ,
\end{align}
where we have used the identity $\Tr([A, B]C) = \Tr([C, A]B)$. Applying the operator $\delta$ to the equations of motion \eqref{EOM_YM}, we see that
\begin{equation}
	\D_\mu\delta F^{\mu\nu} + \bigl[\delta A_\mu, F^{\mu\nu}\bigr] = 0.
\end{equation}
It then follows that on-shell, i.e.~when the equations of motion are satisfied, the contraction of the presymplectic form gives
\begin{equation}
	\If_\V\Omega_\text{YM} [A, \delta A] = \int_\Sigma \dd\Sigma_\mu\,\partial_\nu\Tr\bigl(\delta F^{\mu\nu}\epsilon\bigr).
\end{equation}
As mentioned earlier, if the surface $\Sigma$ has no boundary then the above integral vanishes as expected. In contrast, considering a finite spacetime region with a boundary implies that $\Sigma$ has a boundary which we denote $\partial\Sigma$, and we obtain
\begin{equation}
	\If_\V\Omega_\text{YM} [A, \delta A] = \int_{\partial\Sigma} \dd\sigma_{\mu\nu}\,\Tr\bigl(\delta F^{\mu\nu}\epsilon\bigr) = \delta H_{\rm YM}[\epsilon],
	\label{iOmegaYM}
\end{equation}
where $\dd\sigma_{\mu\nu} := \tfrac{1}{2}\varepsilon_{\mu\nu\rho\sigma}\, \dd x^\rho \wedge \dd x^\sigma$, while $H_{\rm YM}[\epsilon]$ is the boundary Yang--Mills charge associated to the gauge transformation generator $\epsilon$ and defined as
\begin{equation}
	H_\text{YM}[\epsilon] := \int_{\partial\Sigma} \dd\sigma_{\mu\nu}\,\Tr\bigl(F^{\mu\nu}\epsilon\bigr).
	\label{ChargeYM}
\end{equation}
Using equation \eqref{ChargesAlg}, the algebra of the boundary Yang--Mills charges gives
\begin{equation}
	\left\{ H_\text{YM}[\epsilon_1], H_\text{YM}[\epsilon_2] \right\} = \int_{\partial\Sigma} \dd\sigma_{\mu\nu}\,\Tr\bigl([\epsilon_1, \epsilon_2] F^{\mu\nu} \bigr).
	\label{ChargesAlgYM}
\end{equation}
for every $\epsilon_1,\,\epsilon_2$ in ${\G}$.

\subsection{Extended phase space for Yang--Mills}
\label{sec:extendedYM}

Equation \eqref{iOmegaYM} shows that, in the presence of a boundary, the boundary charges associated to the gauge symmetry and generated by the symplectic potential  $\Omega_\text{YM}$ do not vanish. However, this result implies that the presymplectic form has non-vanishing components in the gauge orbits directions  (see \cite{C.W.1987}), which means that $\Omega_\text{YM}$ cannot be obtained as the pullback of a symplectic form on the gauge reduced field space. If one is to require that the field theory under consideration, defined on a spacetime with a compact boundary, can be described by a gauge reduced field space, and that the pullback of the associated symplectic form $\tilde \Omega$ provides the presymplectic form $\Omega$ on the field space at hand, then on-shell the presymplectic form must satisfy:
\begin{equation}
	\If_\V\Omega = 0 ,
	\label{iOmega=0}
\end{equation}
where $\V$ is an arbitrary vector field tangent to a gauge orbit in the field space. 

It was shown by Donnelly and Freidel in \cite{D.F.2016} that one can construct a theory for Yang--Mills where the presymplectic form satisfies \eqref{iOmega=0} in the presence of boundaries. As we present in the following, this is achieved via an extension of the phase space with new variables associated to these boundaries. 	
Similarly to the construction in \cite{D.F.2016}, but without considering field-dependent gauge transformations, we introduce a group valued field $\varphi$ defined on the boundary $\partial\Sigma$, the Yang--Mills edge mode, and which transforms under gauge transformations as
\begin{equation}
	\varphi \to \varphi g^{-1}.
\end{equation}
Consequently (see eq.\eqref{fieldspaceLiederivative}), the interior product of the differential $\delta \varphi$ with a vector field $\V$ tangent to a gauge orbit in the field space is given by
\begin{equation}
	\If_\V\delta\varphi = \varphi\epsilon.
\end{equation}
The symplectic potential associated with the field $\varphi$ is defined as the boundary integral
\begin{equation}
	\Theta_\varphi [A, \delta A, \varphi, \delta \varphi ] := \int_{\partial\Sigma} \dd\sigma_{\mu\nu}\,\Tr\bigl(F^{\mu\nu}\varphi^{-1}\delta\varphi\bigr),
	\label{ThetaphiYM}
\end{equation}
and the corresponding presymplectic form is
\begin{equation}
	\Omega_\varphi [A, \delta A, \varphi, \delta \varphi ] := \delta\Theta_\varphi = \int_{\partial\Sigma} \dd\sigma_{\mu\nu}\,\Tr\bigl(\delta F^{\mu\nu}\varphi^{-1}\delta\varphi - F^{\mu\nu}\varphi^{-1}\delta\varphi\varphi^{-1}\delta\varphi \bigr).
	\label{Omega_v}
\end{equation}
Under a gauge transformation, the combination $\varphi^{-1}\delta\varphi$ transforms homogeneously:
\begin{equation}
	\varphi^{-1}\delta\varphi \to g\varphi^{-1}\delta\varphi g^{-1}.
\end{equation}
It then immediately follows that the presymplectic form \eqref{Omega_v} is gauge invariant. 
Furthermore, we have
\begin{equation}
	\If_\V(\varphi^{-1}\delta\varphi) = \varphi^{-1}\If_\V\delta\varphi = \epsilon.
\end{equation}
Taking the contraction of the presymplectic form $\Omega_\varphi$, we then obtain
\begin{align}
	\If_\V\Omega_\varphi &= \int_{\partial\Sigma} \dd\sigma_{\mu\nu}\,\Tr\Bigl(\bigl(\If_\V\delta F^{\mu\nu}\bigr)\varphi^{-1}\delta\varphi - \delta F^{\mu\nu}\If_\V(\varphi^{-1}\delta\varphi) \notag \\ & \hspace{75pt} - F^{\mu\nu}\If_\V(\varphi^{-1}\delta\varphi)\varphi^{-1}\delta\varphi + F^{\mu\nu}\varphi^{-1}\delta\varphi \If_\V(\varphi^{-1}\delta\varphi) \Bigr) \notag \\
	&= \int_{\partial\Sigma} \dd\sigma_{\mu\nu}\,\Tr\Bigl([F^{\mu\nu}, \epsilon]\varphi^{-1}\delta\varphi - \delta F^{\mu\nu}\epsilon - F^{\mu\nu}[\epsilon, \varphi^{-1}\delta\varphi]\Bigr) \notag \\
	&= -\int_{\partial\Sigma} \dd\sigma_{\mu\nu}\,\Tr\bigl(\delta F^{\mu\nu}\epsilon\bigr),
	\label{iOmega_v}
\end{align}
where the identity $\Tr([A, B]C) = \Tr(A[B, C])$ was used.

Defining the extended presymplectic form of the theory as
\begin{equation}
	\Omega_\text{YM}^\text{T} [A, \delta A, \varphi, \delta \varphi ] := \Omega_\text{YM} [A, \delta A]  + \Omega_\varphi [A, \delta A, \varphi, \delta \varphi ] ,
	\label{OmegaYMT}
\end{equation}
where $\Omega_\text{YM}$ and $\Omega_\varphi$ are defined respectively by \eqref{OmegaYM} and \eqref{Omega_v}, we obtain that
\begin{equation}
	\If_\V\Omega_\text{YM}^\text{T} = 0 ,
\end{equation}
which means that $\Omega_\text{YM}^\text{T}$ satisfies the condition \eqref{iOmega=0}.

Note that, although the field $\varphi$ and the presymplectic form \eqref{Omega_v} are identical to what we find in  \cite{D.F.2016}, the interpretations of the constructions are quite different. In \cite{D.F.2016}, the introduction of the field $\varphi$ on the boundary is necessary in order to ensure gauge invariance of the presymplectic form under field-dependent gauge transformations. If gauge transformations do not depend on the fields, the presymplectic form given by \eqref{OmegaYM} is gauge invariant by itself, whether or not there is a boundary. However, following \cite{C.W.1987}, gauge invariance of the presymplectic form is not sufficient; one must additionally require that the presymplectic form should satisfy eq.\ \eqref{iOmega=0}. The calculations presented above show that the presymplectic form \eqref{OmegaYM} does not satisfy the condition \eqref{iOmega=0} if the surface $\Sigma$ has a boundary; in contrast, this condition \eqref{iOmega=0} is satisfied by the extended presymplectic form defined in \eqref{OmegaYMT}, and first introduced in \cite{D.F.2016}. 

This concludes our treatment of internal gauge symmetries in the theory Yang--Mills on Minkowski spacetime, and the corresponding edge mode extension. Next, we analyze another category of symmetries in Yang--Mills theory on Minkowski spacetime: the global spacetime symmetries.

\subsection{Global spacetime symmetries in Yang--Mills}\label{sec:globYM}    

In addition to gauge symmetries, the Yang--Mills Lagrangian is also invariant under the global Poincar\'e transformations\footnote{Since Yang--Mills theory is massless, it also possesses a conformal symmetry. Although we will not discuss this symmetry here, it can be treated similarly to the Poincar\'e symmetry.}, with parameters being independent of spacetime points. This symmetry is of interest in the present context for two reasons. First, in the case of global spacetime symmetries, there are non-vanishing charges associated with them. Second, these symmetries are a special case of spacetime diffeomorphisms and therefore they serve as a bridge between theories defined on Minkowski space and the generally covariant ones defined on an arbitrary curved Lorentzian manifold, which we are going to discuss in section \ref{sec:gravityeinsteinhilbert}. In what follows, we will do the computations for the Abelian case of Maxwell theory; the generalization to a non-Abelian Yang--Mills theory is straightforward.

Let us consider global translations first. Maxwell theory is described by the action
\begin{equation}
	S_{\text{EM}}[A] = -\frac14 \int_M \dd^4 x \, F^{\mu\nu}F_{\mu\nu} ,
\end{equation}
where the Lagrangian density $\lem = -\tfrac{1}{4}F^{\mu\nu}F_{\mu\nu}$ is invariant under spacetime translations generated by a constant vector $\lambda^\mu$, $x^\mu \mapsto x'{}^\mu = x^\mu + \lambda^\mu$. These translations induce the following field variations:
\begin{align}
	\Delta_\lambda^{\text{T}} A^\mu &= \lambda^\sigma \partial_\sigma A^\mu, \label{Atranslation} \\
	\Delta_\lambda^{\text{T}} F^{\mu\nu} &= \lambda^\sigma \partial_\sigma F^{\mu\nu}. \label{Ftranslation} 
\end{align}
In order to compute the associated charge we need to contract the field space vector generated by the above transformation,
\begin{equation}
	\V_{\lambda}^{\text{T}} = \int_M \dd^4 x \,\Delta_\lambda^{\text{T}} A^{\mu}(x) \frac{\delta}{\delta A^{\mu}(x)},
\end{equation}
with Maxwell version of the presymplectic form \eqref{OmegaYM}
\begin{equation}
	\Omega_{\text{EM}} = - \int_\Sigma \dd \Sigma_\mu \, \delta F^{\mu\nu}\delta A_\nu.
\end{equation}
We get
\begin{align}
	\If_{\V_\lambda^{\text{T}}} \Omega_{\text{EM}} &= -\Sint{\mu} \lambda^{\sigma}\bigl(\partial_\sigma F^{\mu\nu} \delta A_\nu - \delta F^{\mu\nu} \partial_\sigma A_\nu \bigr)\notag \\
	&= \Sint{\mu} \lambda^\sigma \delta \left(F^{\mu\nu}\partial_\sigma A_\nu \right)- \Sint{\mu}\partial_\sigma \left(\lambda^\sigma F^{\mu\nu}\delta A_\nu\right)\notag \\
	&= \Sint{\mu} \lambda^\sigma \bigl(\delta\left(F^{\mu\nu} \partial_\sigma A_\nu \right)-\partial_\sigma \left(F^{\mu\nu}\delta A_\nu\right)\bigr) - \Sint{\mu}\partial_\sigma \bigl(\lambda^\sigma F^{\mu\nu}\delta A_\nu- \lambda^\mu F^{\sigma \nu} \delta A_\nu\bigr)\notag \\
	&= \Sint{\mu}\lambda^\sigma \delta \bigl(F^{\mu\nu}\partial_\sigma A_\nu + \delta_\sigma^\mu \lem \bigr) - 2\DSint{\mu}{\sigma} \lambda^\sigma F^{\mu\nu} \delta A_\nu. \label{iTranslationOmega}
\end{align}
where the equation of motion $\partial_\mu F^{\mu\nu} = 0$ was used. We thus have
\begin{equation}\label{TranslationFlux}
	\If_{\V_{\lambda}^{\text{T}}}\Omega_{\text{EM}} = \delta \tilde{H}_{\text{EM}}^{\rm T}[\lambda]  -2 \DSint{\mu}{\sigma} \lambda^\sigma F^{\mu\nu}\delta A_\nu,
\end{equation}
where the charge $\tilde{H}_{\text{EM}}^{\rm T}[\lambda]$ is given by the integral of the canonical energy-momentum tensor $\tilde T \updown{\mu}{\sigma}$:
\begin{equation}\label{canonicalEMtensor}
	\tilde T \updown{\mu}{\sigma} := F^{\mu\nu}\partial_\sigma A_\nu + \delta^{\mu}_\sigma\lem
\end{equation}
We thus obtained the canonical translation charge with an additional corner term that can not be written as a total field space derivative. This additional term is called a \textit{flux}, and in general it reflects the fact that the canonical energy-momentum tensor is not conserved, or that the system is dissipative (there might be e.g., electromagnetic radiation going through the boundary). The reason it appears here and not in the previously considered gauge charges is because translations also move the location of the boundary. In order to restore integrability of the charge, one needs to take the embedding into account. This is accomplished by the extended phase space formalism, which is covered in section 6.

Furthermore, the charge $\tilde{H}_{\text{EM}}^{\rm T}[\lambda]$ can be rewritten as
\begin{align}
	\tilde{H}_{\text{EM}}^{\rm T}[\lambda] &= \Sint{\mu}\lambda^\sigma \bigl(F^{\mu\nu}\partial_\sigma A_\nu + \delta^\mu_\sigma \lem \bigr)\notag \\
	&= \Sint{\mu}\lambda^\sigma \bigl(F^{\mu\nu}F_{\sigma\nu} + \partial_\nu(F^{\mu\nu}A_\sigma) +\delta^\mu_\sigma \lem \bigr)\notag \\
	&= \Sint{\mu}\lambda^\sigma \bigl(F^{\mu\nu}F_{\sigma\nu} + \delta^\mu_\sigma \lem\bigr) + \DSint{\mu}{\nu} \lambda^\sigma F^{\mu\nu}A_\sigma\notag \\
	&= H_{\text{EM}}^{\rm T}[\lambda] + H_{\text{EM}}[\lambda^\sigma A_\sigma]. \label{TranslationCharge}
\end{align}
We thus obtain the translation charge $H_{\text{EM}}^{\rm T}[\lambda]$ defined\footnote{Note that the translation charge has support in the bulk. This is because global translations are not local transformations and thus, Noether's second theorem does not apply.} as the integral over $\Sigma$ of the symmetric energy-momentum tensor
\begin{equation}
	T \updown{\mu}{\sigma} := F^{\mu\nu}F_{\sigma\nu} + \delta^{\mu}_\sigma\lem
	\label{symmetricEMtensor}
\end{equation}
contracted with $\lambda^\sigma$.

The second term in \eqref{TranslationCharge} corresponds to a gauge charge, as given in \eqref{ChargeYM}, with the caveat that the corresponding gauge transformation is field dependent. The presence of this particular boundary gauge charge reflects the standard issue about the gauge invariance of the canonical energy-momentum tensor \eqref{canonicalEMtensor} in Maxwell theory, and can be understood as a consequence of the fact that the translations given by \eqref{Atranslation}--\eqref{Ftranslation} do not commute with gauge transformations. One could get rid of the gauge charge contribution in \eqref{TranslationCharge} if from the beginning one extends the definition of the transformation \eqref{Atranslation} to include a field dependent gauge transformation in addition to a pure translation (see for instance  \cite{Scheck:2012gam} or \cite{Mieling2017NoethersTA}, where this process is carried out within the conventional framework of Noether's theorem.) The infinitesimal change in $A^\mu$ under such a generalized transformation is defined to be
\begin{equation}
	\tilde\Delta_\lambda^{\rm T}A^\mu = \lambda^\sigma\partial_\sigma A^\mu - \partial^\mu(\lambda^\sigma A_\sigma) = \lambda^\sigma F\downup{\sigma}{\mu},
	\label{Atranslation-generalized}
\end{equation}
while the field tensor still transforms as in \eqref{Ftranslation}. Note that this makes the field variation $\tilde\Delta_\lambda^{\rm T}A^\mu$ independent of the choice of gauge for $A^\mu$. For the extended transformation \eqref{Atranslation-generalized} we have
\begin{equation}
	\If_{\tilde V_\lambda^{\rm T}}\Omega_{\rm EM} = \delta H_{\rm EM}^{\rm T}[\lambda] - 2\int_{\partial\Sigma} \dd\sigma_{\mu\sigma} \lambda^\sigma F^{\mu\nu}\delta A_\nu,
\end{equation}
with the charge $H_{\rm EM}^{\rm T}[\lambda]$ corresponding to the symmetric energy-momentum tensor \eqref{symmetricEMtensor}.

Let us now turn our focus to Lorentz transformations. The potential and field strength transform as
\begin{align}
	\Delta_\lambda^{\text{L}} A_\mu &= \lambda\updown{\rho}{\sigma} x^\sigma \partial_\rho A_\mu + \lambda \updown{\nu}{\mu} A_\nu,\\
	\Delta_\lambda^{\text{L}} F^{\mu\nu} &= \lambda\updown{\rho}{\sigma} x^{\sigma} \partial_\rho F^{\mu\nu} - \lambda \updown{\mu}{\rho}F^{\rho\nu} - \lambda\updown{\nu}{\rho}F^{\mu\rho}, 
\end{align}
where $\lambda^{\mu\nu} = - \lambda^{\nu\mu}$ generates infinitesimal Lorentz transformations.
The contraction of the associated field space vector with the symplectic two-form gives
\begin{align}
	\If_{\V_{\lambda}^{\text{L}}} \Omega_{\text{EM}} &= - \Sint{\mu} \Bigr(\lambda \updown{\rho}{\sigma} x^\sigma \partial_\rho F^{\mu\nu} \delta A_\nu - \lambda \updown{\mu}{\rho}F^{\rho\nu}\delta A_\nu - \lambda \updown{\nu}{\rho}F^{\mu\rho} \delta A_\nu \notag \\
	&\hspace{72pt}- \delta F^{\mu\nu}\lambda \updown{\rho}{\sigma}x^{\sigma}\partial_\rho A_\nu-\delta F^{\mu\nu} \lambda \updown{\rho}{\nu}A_{\rho}\Bigr)\notag \\
	&= \Sint{\mu}\Bigl(\lambda\updown{\rho}{\sigma} \, \delta \bigl(x^\sigma F^{\mu\nu}\partial_\rho A_\nu + F^{\mu\sigma}A_\rho\bigr) + \lambda \updown{\mu}{\rho}F^{\rho\nu}\delta A_\nu -\partial_\rho \left(\lambda \updown{\rho}{\sigma} x^\sigma F^{\mu\nu}\delta A_\nu \right)\Bigr)\notag \\
	&= \Sint{\mu} \Bigl(\lambda\updown{\rho}{\sigma} \, \delta \bigl(x^\sigma F^{\mu\nu}\partial_\rho A_\nu + F^{\mu\sigma}A_\rho\bigr) - \lambda \updown{\mu}{\sigma} x^\sigma F^{\rho\nu}\partial_\rho \delta A_\nu \notag \\
	&\hspace{72pt}- \partial_\rho \left(\lambda \updown{\rho}{\sigma} x^\sigma F^{\mu\nu}\delta A_\nu -\lambda \updown{\mu}{\sigma}x^\sigma F^{\rho\nu}\delta A_\nu \right)\Bigr)
\end{align}
and we therefore have
\begin{equation}\label{LorentzFlux}
	\If_{\V_{\lambda}^{\text{L}}} \Omega_{\text{EM}} =\delta \tilde{H}_{\text{EM}}^{\rm L}[\lambda] -2\DSint{\mu}{\rho} \lambda \updown{\rho}{\sigma}x^{\sigma}F^{\mu\nu}\delta A_\nu ,
\end{equation}
where we again obtain a charge term
\begin{equation}
	\tilde H^{\rm L}_{\rm EM}[\lambda] := \Sint{\mu} \lambda \updown{\rho}{\sigma} \left(F^{\mu\nu}x^\sigma \partial_\rho A_\nu + A_\rho F^{\mu\sigma }+x^{\sigma}\delta^\mu_\rho \lem \right)
	\label{LorentzCharge}
\end{equation}
and a flux term which reflects the fact that the relativistic angular momentum is not conserved, because Lorentz transformations do not preserve the boundary.

The charge \eqref{LorentzCharge} can be written as
\begin{align}\label{LorentzCharge2}
	\tilde{H}^{L}_{\text{EM}}[\lambda] &= \Sint{\mu} \lambda \updown{\rho}{\sigma} \left(x^{\sigma}F^{\mu\nu}F_{\rho\nu} + x^{\sigma}F^{\mu\nu}\partial_\nu A_\rho + A_\rho F^{\mu\sigma} + \delta^\mu_\rho x^{\sigma}\lem \right) \notag \\
	&= \Sint{\mu} \lambda \updown{\rho}{\sigma} \bigl( x^{\sigma}F^{\mu\nu}F_{\rho\nu}+ \delta^\mu_\rho x^{\sigma}\lem\bigr) + \DSint{\mu}{\nu} \lambda\updown{\rho}{\sigma}x^{\sigma}F^{\mu\nu}A_\rho \notag \\
	&= H^{\rm L}_{\text{EM}}[\lambda] + H_{\text{EM}}[\lambda \updown{\rho}{\sigma}x^{\sigma} A_\rho].
\end{align}
After a short calculation, we find that the first term in \eqref{LorentzCharge2} can be expressed as
\begin{equation}
	H_{\text{EM}}^{L}[\lambda] = \frac12 \Sint{\mu} \lambda_{\rho\sigma}M^{\sigma\mu\rho},
\end{equation}
where
\begin{equation}
	M^{\sigma\rho\mu} := x^{\sigma}T^{\mu\rho}- x^{\rho}T^{\mu\sigma}.
\end{equation}
is the angular momentum tensor. 

Finally, similarly to the case of translations, the second term in \eqref{LorentzCharge2} is a gauge charge \eqref{ChargeYM} corresponding to a field dependent gauge transformation. This can again be seen as a consequence of the fact that Lorentz transformations do not commute with gauge transformations of the field $A_\mu$. This concludes this section about global spacetime symmetries for Yang--Mills theory, and we now move to the treatment of Einstein--Hilbert formulation of the gravity theory.

\section{Gravity: Einstein--Hilbert action}\label{sec:gravityeinsteinhilbert}

General relativity in the Einstein--Hilbert metric formulation is defined by the action
\begin{equation}
	S_\text{EH} [g] = \int \dd^4x L \, := \frac{1}{2\kappa} \int \dd^4x\,\sqrt{-g}g^{\mu\nu}R_{\mu\nu} ,
\end{equation}
where $L$ is the Einstein--Hilbert Lagrangian scalar density, $\kappa = 8 \pi G$ is from now set equal to $1$, $g$ is the determinant of the Lorentzian metric $g_{\mu\nu}$, $g^{\mu \nu}$ is the inverse metric, and $R_{\mu\nu}$ is the Ricci tensor given by
\begin{equation}
	R_{\mu\nu} = \partial_\alpha \Gamma^\alpha_{\mu\nu} - \partial_\mu \Gamma^\alpha_{\nu\alpha} + \Gamma^\alpha_{\mu\nu}\Gamma^\beta_{\alpha\beta} - \Gamma^\alpha_{\mu\beta}\Gamma^\beta_{\nu\alpha} ,
\end{equation}
with
\begin{equation}
	\Gamma^\alpha_{\mu\nu} = \frac{1}{2}g^{\alpha\beta}\bigl(\partial_\mu g_{\beta\nu} + \partial_\nu g_{\mu\beta} - \partial_\beta g_{\mu\nu}\bigr)
\end{equation}
being the connection compatible with the metric: $\nabla_\alpha g_{\mu\nu} := \partial_\alpha g_{\mu\nu} - \Gamma^\beta_{\alpha\mu} g_{\beta\nu} - \Gamma^\beta_{\alpha\nu} g_{\beta\mu} = 0$. 

The variation of the action with respect to the metric yields
\begin{equation}
	\delta S_\text{EH} [g, \delta g] = \frac{1}{2}\int \dd^4x\,\biggl[\sqrt{-g}\biggl(R_{\mu\nu} - \frac{1}{2}Rg_{\mu\nu}\biggr)\delta g^{\mu\nu} + \sqrt{-g}g^{\mu\nu}\delta R_{\mu\nu}\biggr] ,
	\label{dS_GR1}
\end{equation}
where we have
\begin{align}
	\delta g^{\mu\nu} &= - g^{\mu\alpha} g^{\nu\beta} \delta g_{\alpha \beta} , \label{dg-1}\\[1ex]
	\delta R_{\mu\nu} &= \nabla_\alpha\delta\Gamma^\alpha_{\mu\nu} - \nabla_\mu\delta\Gamma^\alpha_{\nu\alpha} , \label{dR} \\[1ex]
	\delta\Gamma^\alpha_{\mu\nu} &= \frac{1}{2}g^{\alpha\beta}\Bigl(\nabla_\mu\delta g_{\beta\nu} + \nabla_\nu\delta g_{\mu\beta} - \nabla_\beta\delta g_{\mu\nu}\Bigr) . \label{dG}
\end{align}
Note that the second term in \eqref{dR} is symmetric in $\mu$ and $\nu$ even if this is not immediately apparent.

\subsection{Presymplectic form and its properties}

The boundary term in the variation of the action arises from the last term in \eqref{dS_GR1}, as one can express the term $g^{\mu\nu}\delta R_{\mu\nu}$ as a total derivative, namely
\begin{equation}
	g^{\mu\nu}\delta R_{\mu\nu} = \nabla_\alpha\bigl(g^{\mu\nu}\delta\Gamma^\alpha_{\mu\nu}\bigr) - \nabla_\mu\bigl(g^{\mu\nu}\delta\Gamma^\alpha_{\alpha\nu}\bigr) = \nabla_\mu\Bigl(g^{\alpha\beta}\delta\Gamma^\mu_{\alpha\beta} - g^{\mu\alpha}\delta\Gamma^\beta_{\alpha\beta}\Bigr).
\end{equation}
Therefore the variation of the Einstein--Hilbert action has the form
\begin{equation}
	\delta S_\text{EH} [g, \delta g] = \frac{1}{2} \int \dd^4x\,\biggl[\sqrt{-g}\biggl(R_{\mu\nu} - \frac{1}{2}Rg_{\mu\nu}\biggr)\delta g^{\mu\nu} + \sqrt{-g}\nabla_\mu v^\mu\biggr]
	\label{dS_GR2}
\end{equation}
where the vector $v^\mu$ is given by
\begin{equation}
	v^\mu := g^{\alpha\beta}\delta\Gamma^\mu_{\alpha\beta} - g^{\mu\alpha}\delta\Gamma^\beta_{\alpha\beta}
	= g^{\mu\alpha}\nabla^\beta\delta g_{\alpha\beta} - g^{\alpha\beta}\nabla^\mu\delta g_{\alpha\beta}.
	\label{v}
\end{equation}
By applying the identity
\begin{equation}
	\nabla_\mu v^\mu = \frac{1}{\sqrt{-g}}\partial_\mu\bigl(\sqrt{-g} v^\mu\bigr) ,
\end{equation}
the boundary term in the variation $\delta S_\text{EH}$ can be written as
\begin{align}
	\frac{1}{2}\int \dd^4x\,\sqrt{-g}g^{\mu\nu}\delta R_{\mu\nu} = \frac{1}{2}\int \dd^4x\,\sqrt{-g}\nabla_\mu v^\mu = \frac{1}{2}\int_{\partial M} \dd\Sigma_\mu\,\sqrt{-g} v^\mu
\end{align}
where the last integral is taken over the boundary $\partial M$ of the spacetime manifold $M$, with $\dd\Sigma_\mu := \tfrac{1}{3!}\varepsilon_{\mu\nu\rho\sigma}\, \dd x^\nu \wedge \dd x^\rho \wedge \dd x^\sigma$, $\varepsilon$ being the Levi-Civita tensor density.

The symplectic potential is then obtained by taking the integrand of this boundary integral and integrating it over a submanifold $\Sigma$ of a Cauchy surface:
\begin{equation}
	\Theta_\text{EH} [g, \delta g] := \int_\Sigma \dd\Sigma_\mu\,\theta_\text{EH}^\mu [g, \delta g] ,
	\label{ThetaEH}
\end{equation}
where the symplectic potential current is
\begin{align}
	\theta_\text{EH}^\mu [g, \delta g] :=& \frac{1}{2}\sqrt{-g}\Bigr(g^{\alpha\beta}\delta\Gamma^\mu_{\alpha\beta} - g^{\mu\alpha}\delta\Gamma^\beta_{\alpha\beta}\Bigr) = \frac{1}{2} \sqrt{-g}\Bigl(g^{\mu\beta}\nabla^\alpha\delta g_{\alpha\beta} - g^{\alpha\beta}\nabla^\mu\delta g_{\alpha\beta} \Bigr) .
	\label{thetaEH}
\end{align}
The presymplectic form is then defined as
\begin{equation}
	\Omega_\text{EH} [g, \delta g] := \delta\Theta_\text{EH}[g, \delta g] .
\end{equation}
Taking the variation of \eqref{ThetaEH}, we find
\begin{equation}
	\Omega_\text{EH} [g, \delta g] = \int_\Sigma \dd\Sigma_\mu\,\omega_\text{EH}^\mu
	\label{OmegaEH}
\end{equation}
with the symplectic current given by
\begin{equation}
	\omega_\text{EH}^\mu [g, \delta g] := \frac{1}{2}\left(\delta\bigl(\sqrt{-g}g^{\alpha\beta}\bigr)\delta\Gamma^\mu_{\alpha\beta} - \delta\bigl(\sqrt{-g}g^{\mu\alpha}\bigr)\delta\Gamma^\beta_{\alpha\beta} \right).
	\label{omegaEH1}
\end{equation}
Thanks to the fact that $\delta g_{\alpha\beta}$ and $\delta\Gamma^\mu_{\alpha\beta}$ transform as tensors under the action of diffeomorphisms, the symplectic current $\omega_\text{EH}^\mu$ transforms as a vector density. Consequently, the presymplectic form $\Omega_\text{EH}$, being an integral of a scalar density, is invariant under the action of diffeomorphisms which preserve the boundary of $\Sigma$, and it transforms in a covariant way under the action of a general diffeomorphism.

\subsection{Gravitational charges and their algebra}

As we established, the presymplectic form $\Omega_\text{EH}$ transforms in a covariant way under the action of diffeomorphisms, and is invariant under the action of boundary preserving diffeomorphisms. Now, as in the Yang--Mills case, we would like to first derive the gravity boundary charges by computing the contraction $\If_\V\Omega_\text{EH}$, where $\V$ is the tangent vector field of a gauge orbit of diffeomorphisms in the field space, and we perform the calculations on-shell.

Under an infinitesimal diffeomorphism generated by a spacetime vector field $\epsilon^\mu$, the variation of the metric is given by
\begin{equation}
	\Delta_\epsilon g_{\mu\nu} = {\cal L}_\epsilon g_{\mu \nu} = \nabla_\mu\epsilon_\nu + \nabla_\nu\epsilon_\mu
\end{equation}
while the variation of the inverse metric is
\begin{equation}
	\Delta_\epsilon g^{\mu\nu} = -g^{\mu\alpha}g^{\nu\beta} {\cal L}_\epsilon g_{\alpha\beta} = -\nabla^\mu\epsilon^\nu - \nabla^\nu\epsilon^\mu.
\end{equation}
where ${\cal L}$ is the spacetime Lie derivative.
Denoting by $\V$ the field space tangent vector field corresponding to the infinitesimal transformation, it follows from Cartan's formula that
\begin{align}
	\If_\V\delta g_{\mu\nu} &= \nabla_\mu\epsilon_\nu + \nabla_\nu\epsilon_\mu, \\[1ex]
	\If_\V\delta g^{\mu\nu} &= -\nabla^\mu\epsilon^\nu - \nabla^\nu\epsilon^\mu.
\end{align}
where $\If_\V$ stands again for the interior product.

Consequently, the interior product of the symplectic potential current $\theta^\mu_{\rm EH}$ in \eqref{thetaEH} gives
\begin{equation}
	\If_\V\theta_\text{EH}^\mu [g] = \frac{1}{2}\nabla_\nu\Bigl(\sqrt{-g}\bigl(\nabla^\nu\epsilon^\mu - \nabla^\mu\epsilon^\nu\bigr)\Bigr) = -\nabla_\nu\xi^{\mu\nu},
	\label{ithetaEH}
\end{equation}
where
\begin{equation}
	\xi^{\mu\nu} := \frac{1}{2}\sqrt{-g}\bigl(\nabla^\mu\epsilon^\nu - \nabla^\nu\epsilon^\mu\bigr).
	\label{xi}
\end{equation}
A more involved calculation (see appendix \ref{sec:gravity_presymplectic_current} for details)  shows that the interior product of the symplectic current $\omega_\text{EH}^\mu$ can be expressed in terms of $\theta^\mu_{\rm EH}$ and $\xi^{\mu\nu}$ as
\begin{equation}
	\If_\V\omega_\text{EH}^\mu [g, \delta g] = \partial_\nu\Bigl(-\epsilon^\mu\theta_\text{EH}^\nu + \epsilon^\nu\theta_\text{EH}^\mu + \delta\xi^{\mu\nu}\Bigr).
	\label{iomegaEH1}
\end{equation}
Consider then the interior product of the presymplectic form,
\begin{align}
	\If_\V\Omega_\text{EH} &= \int_\Sigma \dd\Sigma_\mu\,\If_\V\omega_\text{EH}^\mu.
	\label{iOmega_EH}
\end{align}	
Introducing the functional
\begin{equation}
	H_\text{EH} [\epsilon]:=  \int_{\partial\Sigma} \dd\sigma_{\mu\nu}\,\sqrt{-g}\ \nabla^\mu\epsilon^\nu ,
	\label{ChargeEH}
\end{equation}
where $\dd\sigma_{\mu\nu} := \tfrac{1}{2}\varepsilon_{\mu\nu\rho\sigma}\, \dd x^\rho \wedge \dd x^\sigma$, we see from \eqref{iomegaEH1} that the interior product can be written in the form
\begin{equation}
	\If_\V\Omega_\text{EH} [g, \delta g] = \delta H_\text{EH}[\epsilon] - 2\int_{\partial\Sigma} \dd\sigma_{\mu\nu}\, \epsilon^\mu \theta_\text{EH}^\nu.
	\label{iOmegaEH}
\end{equation}

In the presence of a boundary, the presymplectic form has non vanishing components along the gauge orbits generated by the diffeomorphisms. A simplification occurs when one considers only diffeomorphisms which preserve the boundary, \ie the corresponding spacetime vector field $\epsilon_{\shortparallel}^\mu$ is tangential to the boundary. Namely one obtains $\dd\sigma_{\mu\nu}\epsilon_{\shortparallel}^\mu = 0$, and therefore for such diffeomorphisms the integral in \eqref{iOmegaEH} vanishes and we have
\begin{equation}
	\If_{\V_{\shortparallel}}\Omega_\text{EH} = \delta H_\text{EH}[\epsilon_{\shortparallel}] .
\end{equation}
The functional $H_\text{EH}[\epsilon_{\shortparallel}]$, known as the Komar charge, corresponds to the gravitational boundary charge associated to the gauge transformation generator $\epsilon_{\shortparallel}$. Following \eqref{ChargesAlg}, and using \eqref{ithetaEH}, the gravitational charges satisfy the algebra
\begin{equation}
	\left\{ H_\text{EH}[v_{\shortparallel}], H_\text{EH}[w_{\shortparallel}] \right\} = \int_{\partial\Sigma} \dd\sigma_{\mu\nu}\,\sqrt{-g}\, \nabla^\mu \bigl(w_{\shortparallel}^\alpha \nabla_\alpha v_{\shortparallel}^\nu - v_{\shortparallel}^\alpha \nabla_\alpha w_{\shortparallel}^\nu\bigr) = - H_\text{EH}\bigl[[v_{\shortparallel},w_{\shortparallel}]\bigr].
	\label{ChargesAlgEH}
\end{equation}

Similarly to the Yang--Mills case discussed in section \ref{sec:extendedYM}, equation \eqref{iOmegaEH} shows that the boundary charges associated to the diffeomorphism symmetry and generated by the symplectic potential $\Omega_\text{EH}$ do not vanish. However, unlike Yang--Mills, there is an important difference arising in gravity which is that $\If_\V\Omega_\text{EH}$ does not correspond to a total field differential. There is an additional term emanating from the fact that diffeomorphisms do not preserve the boundary in general. As discussed earlier, restricting to boundary preserving diffeomorphisms leads to the identification of boundary charges and derive their algebra. Nevertheless, one would like to be able to treat all diffeomorphisms on the same footing and to define an algebra of charges associated to the entire group of symmetry of the theory.  In \cite{Freidel:2021cjp,Ciambelli:2021nmv}, it has been shown that one can realize this outcome by considering the embedding map into spacetime as independent variables in the theory, and consequently introducing an extended phase space for the theory. We present the construction of this extension and its consequence in the following section. The embedding map will eventually define the so-called gravitational edge mode.

Going back to the original question of a non-vanishing $\If_\V\Omega_\text{EH}$, the discussion does not differ from the case of Yang--Mills. Namely, the condition $\If_\V\Omega_\text{EH} \neq 0$ implies that the presymplectic form has non-vanishing components in the gauge orbits directions  (see \cite{C.W.1987}), and as a consequence the presymplectic form $\Omega_\text{EH}$ in \eqref{OmegaEH} cannot be obtained as the pullback of a symplectic form on the gauge reduced field space. If one is to have such requirement, then the  presymplectic form $\Omega_\text{EH}$ has to be modified appropriately. We will indeed show in section \ref{sec:gravity_boundary_symplectic_form} that, in the context of the extended phase space for gravity, one is able to impose the condition above by introducing an edge mode boundary term in the definition of the presymplectic form.

\subsection{Examples of boundary charges for the Kerr--Newman--de Sitter spacetime}

In this section we provide an example of evaluating the boundary charges
\begin{equation}
	H_{\rm EH}[\epsilon] = \int_S \dd\sigma_{\mu\nu}\,\sqrt{-g}\nabla^\mu\epsilon^\nu
\end{equation}
for a particular solution of the Einstein equations. We consider the Kerr--Newman--de Sitter spacetime, which is described by the metric
\begin{align}
	ds^2 = {}&\frac{1}{\Xi^2\rho^2}\Bigl(-\Delta_r + a^2\Delta_\theta\sin^2\theta\Bigr)dt^2 + \frac{\rho^2}{\Delta_r}dr^2 + \frac{\rho^2}{\Delta_\theta}d\theta^2 \notag \\
	&+ \frac{1}{\Xi^2\rho^2}\Bigl(\bigl(r^2 + a^2\bigr)^2\Delta_\theta - a^2\Delta_r\sin^2\theta\Bigr)\sin^2\theta\,d\phi^2 \notag \\
	&+ \frac{2a}{\Xi^2\rho^2}\Bigl(\Delta_r - \bigl(r^2 + a^2\bigr)\Delta_\theta\Bigr)\sin^2\theta\,dt\,d\phi
	\label{ds2}
\end{align}
where
\begin{align}
	\rho^2 &= r^2 + a^2\cos^2\theta \\
	\Xi &= 1 + \frac{\Lambda a^2}{3} \\
	\Delta_r &= \biggl(1 - \frac{\Lambda}{3}r^2\biggr)\bigl(r^2 + a^2\bigr) - 2Mr + Q^2 \\
	\Delta_\theta &= 1 + \frac{\Lambda a^2}{3}\cos^2\theta.
\end{align}
The surface $S$ is taken to be a coordinate sphere at a finite value $r=R$ of the radial coordinate. 

Note that the Kerr-Newman-de Sitter spacetime is a solution of Einstein-Maxwell equations with non-vanishing cosmological constant and energy-momentum tensor for a point charge, however these extensions do not alter the expression for the diffeomorphism boundary charge. This is due to the fact that both the cosmological constant term and the Maxwell term in the Lagrangian do not contain derivatives of the metric, so the variations of these terms with respect to the metric do not produce total derivatives and therefore they do not contribute to the boundary term. We can therefore proceed with the calculation of the diffeomorphism boundary charge without any modification.

For the Kerr-Newman-de Sitter spacetime, the integral to be computed takes the form
\begin{equation}
	H_{\rm EH}[\epsilon] = \int_S d\theta\wedge d\phi\,\sqrt{-g}\,\xi[\epsilon]\Big|_{r=R}
	\label{H_sphere}
\end{equation}
where we have introduced the abbreviation
\begin{equation}
	\xi[\epsilon] = \nabla^t\epsilon^r - \nabla^r\epsilon^t
\end{equation}
for the integrand, and the volume element for the metric \eqref{ds2} is
\begin{equation}
	\sqrt{-g} = \frac{\rho^2\,\sin\theta}{\Xi^2}.
\end{equation}
Consider first the charge associated with the timelike Killing vector $\partial_t$. To evaluate the integral \eqref{H_sphere}, we compute the Christoffel symbols of the metric \eqref{ds2} and the components of the tensor $\nabla^\mu\epsilon^\nu$, where now $\epsilon = \partial_t$. We find that the integrand $\xi[\partial_t]$ is given as an explicit function of the spacetime coordinates by
\begin{equation}
	\xi[\partial_t] = 2\frac{\bigl(r^2 + a^2\bigr)}{\rho^2}\biggl[\frac{-M\bigl(2r^2 - \rho^2\bigr) + Q^2r}{\rho^4} + \frac{\Lambda}{3}r\biggr].
\end{equation}
Then integration over the sphere $S$ yields the charge
\begin{equation}
	H_{\rm EH}[\partial_t] = \frac{8\pi}{\Xi^2}\biggl[-M + \frac{Q^2}{2R} + \frac{Q^2}{2aR^2}\bigl(R^2 + a^2\bigr)\arctan\biggl(\frac{a}{R}\biggr) + \frac{\Lambda}{3}R\bigl(R^2 + a^2\bigr)\biggr].
\end{equation}
A non-trivial charge is also associated with the spacelike Killing vector $\partial_\phi$. In this case we have the integrand
\begin{equation}
	\xi[\partial_\phi] = \frac{2a \sin^2\theta}{\rho^6}\biggl[ M\Bigl(2r^2\bigl(r^2 + a^2\bigr) + \rho^2\bigl(r^2 - a^2\bigr)\Bigr) - Q^2r\bigl(r^2 + \rho^2 + a^2\bigr)\biggr]
\end{equation}
and the corresponding charge is given by
\begin{equation}
	H_{\rm EH}[\partial_\phi] = \frac{16\pi}{\Xi^2}\biggl[Ma + \frac{Q^2}{4}\biggl(\frac{R}{a} - \frac{a}{R}\biggr) - \frac{Q^2}{4}\biggl(\frac{R}{a} + \frac{a}{R}\biggr)^2\arctan\biggl(\frac{a}{R}\biggr)\biggr].
\end{equation}
The charges corresponding to the remaining coordinate basis vectors, $H_{\rm EH}[\partial_r]$ and $H_{\rm EH}[\partial_\theta]$, are identically vanishing.

To examine the behavior of the charges $H_{\rm EH}[\partial_t]$ and $H_{\rm EH}[\partial_\phi]$ in the limiting cases where $a$, $Q$ or $\Lambda$ go to zero, it is useful to note that when $R \gg a$, the above expressions for these charges have the forms
\begin{align}
	H_{\rm EH}[\partial_t] &= \frac{8\pi}{\Xi^2}\biggl[-M + \frac{\Lambda}{3}R^3\left( 1 + \frac{a^2}{R^2} \right) + \frac{Q^2}{R} \left( 1 + \frac{a^2}{3R^2} + {\cal O}\left(\frac{a^4}{R^4} \right) \right) \biggr]\\
	H_{\rm EH}[\partial_\phi] &= \frac{16\pi}{\Xi^2}\Biggl[Ma - Q^2 \left( \frac{2a}{3R} + {\cal O}\left(\frac{a^3}{R^3}\right) \right)\Biggr]
\end{align}
The results of some of the limiting cases corresponding to well known spacetimes are summarized in the following table:

\bgroup
\def\arraystretch{2}
\begin{center}
	\begin{tabular}[h]{|c|c|c|c|}
		\hline
		\textbf{Spacetime} & \textbf{Limit} & \textbf{Charge} $H_{\rm EH}[\partial_t]$ & \textbf{Charge} $H_{\rm EH}[\partial_\phi]$ \\
		\hline
		Schwarzschild & $a, Q , \Lambda \to 0$ & $-8\pi M$ & $0$ \\
		\hline
		Reissner-Nordstr\"om & $a, \Lambda \to 0$ & $8\pi\biggl(-M + \dfrac{Q^2}{R} \biggr)$ & $0$ \\
		\hline
		Kerr & $Q , \Lambda \to 0$ & $-8\pi M$ & $16\pi Ma$ \\
		\hline
		de Sitter & $a, Q \to 0$ & $8\pi\biggl(-M + \dfrac{\Lambda}{3}R^3 \biggr)$ & $0$ \\
		\hline
	\end{tabular}
\end{center}
\egroup

\section{Extended phase space for theories with spacetime symmetries} \label{sec:extendedphasespace}

In the context of theories with spacetime symmetries, one could introduce an additional independent field associated to the boundary, which would provide counter-terms to the flux terms obtained when computing the contraction of the presymplectic form for spacetime symmetries, as illustrated in \eqref{TranslationFlux} and \eqref{LorentzFlux} for Yang--Mills theory, and in \eqref{iOmegaEH} for pure gravity. This boundary field emerges from the introduction of the spacetime embedding map as an independent phase space variable, and it constitutes the so called gravitational edge mode. In the following, we show how this edge mode is obtained and derive its contribution to the presymplectic form.

\subsection{Spacetime embedding as a new phase space variable}
\label{sec:embedding}

Consider an embedding map $\phi$, which maps its domain $\D \subset \R^4$ onto the spacetime $M$, and under which a closed subspace ${\cal R} \subset \D$ is mapped to a subregion $R \subset M$. For clarity, we denote points in the domain $\D$ as $\X, \Y, \dots$ while spacetime points are denoted by $x, y, \dots$ as usual. Thus,
\begin{equation}
	\begin{aligned}
		&\phi : \D \subset \R^4 \rightarrow M \\
		&\phi(\X) = x
	\end{aligned}
\end{equation}
After introducing the embedding field $\phi$, the action of a generally covariant theory, for some field variables $\varphi^a$, associated to the spacetime region $R$ takes the form
\begin{equation}
	S_R [\varphi^a, \phi]= \int_{\cal R} \phi^* {\bf L} = \int_{\cal R} \dd^4\X\ J(\X) L\bigl(\phi(\X)\bigr),
	\label{S_R}
\end{equation}
where ${\bf L}(x):= L(x)\,\dd^4x$, $L$ is the Lagrangian scalar density of the theory, and $J$ is the Jacobian:
\begin{equation}
	J(\X) = \det\biggl(\frac{\partial \phi^\mu(\X)}{\partial \X^\nu}\biggr).
\end{equation}
We assume that the embedding $\phi$ is a new independent phase space variable in the theory. As such, and in order to perform the calculations on the field space, we rely on the machinery of the covariant phase space formalism which was briefly summarized in section \ref{sec:covariant_phase_space}.

A central result concerning the embedding map $\phi$ has to do with the variation (field space differential) of an embedded functional, such as the action integral \eqref{S_R}. More generally, one may wish to find the variation of an integral of the form
\begin{equation}
	{\cal I} := \int_{\cal S} \dd\Sigma_{\mu_1\cdots\mu_n}\bigl(\phi(\X)\bigr) \alpha^{\mu_1\cdots\mu_n}\bigl(\phi(\X)\bigr)
	\label{I}
\end{equation}
where ${\cal S} \subset {\cal D}$ is a submanifold of codimension $n$, $\alpha^{\mu_1\cdots\mu_n}$ is a tensor density of rank $n$, and the volume form $\dd\Sigma_{\mu_1\dots\mu_n}\bigl(\phi(\X)\bigr)$ arises via pullback from the natural volume form $\dd\Sigma_{\mu_1\cdots\mu_n}(x) = \frac{1}{(d-n)!}\epsilon_{\mu_1\cdots\mu_n\mu_{n+1}\cdots\mu_d}\dd x^{\mu_{n+1}}\wedge\cdots\wedge \dd x^{\mu_d}$ on the spacetime submanifold $S := \phi^{-1}({\cal S}) \subset M$ (here $d=4$ is the spacetime dimension). The variation of the integral \eqref{I} is computed in Appendix \ref{sec:embedded_variation}. To concisely express the contribution arising from the variation of the embedding map, it is useful to introduce the variational vector field
\begin{equation}
	\chi^\mu(x) := \bigl(\delta \phi^\mu \circ \phi^{-1}\bigr)(x)
	\label{chi}
\end{equation}
which is a one-form on field space and transforms as a vector field under spacetime diffeomorphisms. The calculation carried out in Appendix \ref{sec:embedded_variation} then shows that the variation $\delta{\cal I}$ is given by
\begin{equation}
	\delta{\cal I} = \int_{\cal S} \dd\Sigma_{\mu_1\cdots\mu_n}\bigl(\phi(\X)\bigr)\,\Bigl[(\delta\alpha)^{\mu_1\cdots\mu_n}\bigl(\phi(\X)\bigr) + {\cal L}_\chi\alpha^{\mu_1\cdots\mu_n}\bigl(\phi(\X)\bigr)\Bigr].
	\label{dI}
\end{equation}
Thus, the variation of the embedding map gives rise to a term which is the Lie derivative of the tensor density $\alpha^{\mu_1\cdots\mu_d}$ along the vector field $\chi^\mu$.

Using the identity \eqref{dI}, we can now deduce the variation of the action \eqref{S_R}. In general, the variation of the Lagrangian density takes the form
\begin{equation}
	\delta L[\varphi^a] = E_a \delta \varphi^a + \dd \theta[\varphi^a,\delta \varphi^a],
\end{equation}
as given in \eqref{VariationLagrangian}, hence we find
\begin{equation}
	\delta S_R = \int_{\cal R} \dd^4\phi(\X)\, \Bigl[E_{a}\bigl(\phi(\X)\bigr)\delta \varphi^a\bigl(\phi(\X)\bigr) + \partial_\mu\theta^\mu\bigl(\phi(\X)\bigr) + {\cal L}_\chi L\bigl(\phi(\X)\bigr) \Bigr],
	\label{dS_R-result}
\end{equation}
where $\partial_\mu$ denotes differentiation with respect to $x^\mu = \phi^\mu(\X)$.
Here the Lie derivative of the scalar density $L$ is actually a total divergence:
\begin{equation}
	{\cal L}_\chi L = \chi^\mu\partial_\mu L + \bigl(\partial_\mu\chi^\mu\bigr)L = \partial_\mu\bigl(\chi^\mu L\bigr).
\end{equation}
Thus, this term combines in \eqref{dS_R-result} with the symplectic potential current of the non-extended theory to give the extended symplectic potential
\begin{equation}
	\Theta_{\rm ext} = \int_{\cal S} \dd\Sigma_\mu\bigl(\phi(\X)\bigr)\,\Bigl[\theta^\mu\bigl(\phi(\X)\bigr) + \chi^\mu L\bigl(\phi(\X)\bigr)\Bigr],
	\label{Theta_ext}
\end{equation}
where $\cal S \subset \cal R$ is such that $\phi({\cal S}) = \Sigma$, which is a submanifold of a Cauchy surface in $R$.

The next step is to take the variation of the symplectic potential \eqref{Theta_ext} to obtain the extended presymplectic form $\Omega_{\rm ext} = \delta\Theta_{\rm ext}$. Let us begin by establishing an expression for the variation of the vector field $\chi^\mu$. Writing the definition \eqref{chi} in the form
\begin{equation}
	\delta\phi^\mu(\X) = \chi^\mu\bigl(\phi(\X)\bigr)
\end{equation}
and taking the variation of this equation, we obtain
\begin{equation}
	0 = \bigl(\delta\chi^\mu\bigr)\bigl(\phi(\X)\bigr) + \chi^\nu\bigl(\phi(\X)\bigr)\partial_\nu\chi^\mu\bigl(\phi(\X)\bigr)
\end{equation}
from which we see that
\begin{equation}
	\delta\chi^\mu = -\chi^\nu\partial_\nu\chi^\mu.
	\label{deltachi}
\end{equation}
Now, introducing the notation
\begin{equation}
	\theta^\mu_{\rm ext} = \theta^\mu + \chi^\mu L
\end{equation}
and using again the key identity \eqref{dI}, the variation of \eqref{Theta_ext} gives
\begin{equation}
	\Omega_{\rm ext} = \int_{\cal S} \dd\Sigma_\mu\bigl(\phi(\X)\bigr)\,\Bigl[\delta\theta^\mu_{\rm ext}\bigl(\phi(\X)\bigr) + {\cal L}_\chi\theta^\mu_{\rm ext}\bigl(\phi(\X)\bigr)\Bigr].
	\label{Omega_ext}
\end{equation}
Recalling that the non-extended symplectic potential current satisfies $\partial_\mu\theta^\mu = \delta L$ on shell, the Lie derivative in the above equation can be written as
\begin{align}
	{\cal L}_\chi\theta^\mu_{\rm ext} &= \chi^\nu\partial_\nu\theta^\mu_{\rm ext} - \bigl(\partial_\nu\chi^\mu\bigr)\theta^\mu_{\rm ext} + \bigl(\partial_\nu\chi^\nu\bigr)\theta^\mu_{\rm ext} \notag \\
	&= \chi^\mu\partial_\nu\theta^\nu_{\rm ext} + \partial_\nu\bigl(-\chi^\mu\theta^\nu_{\rm ext} - \chi^\nu\theta^\mu_{\rm ext}\bigr) \notag \\
	&= \chi^\mu\delta L + \chi^\mu\partial_\nu\bigl(\chi^\nu L\bigr) + \partial_\nu\bigl(-\chi^\mu\theta^\nu_{\rm ext} - \chi^\nu\theta^\mu_{\rm ext}\bigr).
\end{align}
Inserting this and
\begin{equation}
	\delta\theta^\mu_{\rm ext} = \delta\theta^\mu - \chi^\nu\bigl(\partial_\nu\chi^\mu\bigr)L - \chi^\mu\delta L
\end{equation}
into \eqref{Omega_ext}, we conclude that on-shell the extended presymplectic form can be expressed as a sum of two terms, which are associated respectively with the surface ${\cal S}$ and its boundary $\partial{\cal S}$ which satisfies $\phi(\partial{\cal S}) = \partial \Sigma$, namely:
\begin{equation}
	\Omega_{\rm ext} = \int_{\cal S} \dd\Sigma_\mu\bigl(\phi(\X)\bigr)\,\omega^\mu\bigl(\phi(\X)\bigr) - \int_{\partial{\cal S}} \dd\sigma_{\mu\nu}\bigl(\phi(\X)\bigr)\,\Bigl(\chi^\mu\theta^\nu - \chi^\nu\theta^\mu + \chi^\mu\chi^\nu L\Bigr)\bigl(\phi(\X)\bigr),
	\label{Extended_Omega}
\end{equation}
where $\omega^\mu = \delta\theta^\mu$ denotes the non-extended presymplectic current. Note that the field $\phi$ enters the bulk term only through the embedding action, and appears explicitly, via the variational vector field $\chi^\mu$, in the corner term only.

Note also that in all the above calculations, no assumptions regarding the explicit form of the Lagrangian density $L$ were used. 

\subsection{Charges of global spacetime symmetries in extended Yang--Mills theory}
\label{sec:embedding4}

What makes the global spacetime symmetries of Yang--Mills theory particularly interesting is that they do not preserve finite spacetime regions, in particular the boundary. It is for this reason that we obtain the flux terms in \eqref{TranslationFlux} and \eqref{LorentzFlux}. 
In order to solely produce the charges associated with Poincar\'e symmetry, we need a boundary contribution related to the change of the boundaries. The way to do this issue in the covariant phase space formalism is to introduce a new field $\phi^\mu$ as follows.

Following the formulation and calculations presented in the previous section \ref{sec:embedding}, the extended Maxwell\footnote{The extended theory here is different from the edge mode extension discussed in section \ref{sec:extendedYM}. We will further comment on this at the end of this section.} action is
\begin{align}\label{Actionphi}
	S_{\text{EM}}[A,\phi] = -\frac14\int_\mathcal{R} \dd^4\X\, J(\X)\, F^{\mu\nu}\bigl(\phi(\X)\bigr)F_{\mu\nu}\bigl(\phi(\X)\bigr),
\end{align}
and the resulting (on-shell) extended presymplectic form \eqref{Extended_Omega} gives
\begin{align}\label{extsymplformYM}
	\Omega^{\rm ext}_{\text{EM}}[A, \delta A, \phi, \delta \phi] = &-\int_{\cal S} \dd\Sigma_\mu\bigl(\phi(\X)\bigr)\,\delta F^{\mu\nu} \delta A_\nu  \notag \\
	&+ \int_{\partial{\cal S}} \dd\sigma_{\mu\nu}\bigl(\phi(\X)\bigr)\,\bigl(2\chi^\mu F^{\nu\rho}\delta A_\rho - \chi^\mu\chi^\nu L_{\rm EM} \bigr).
\end{align}
(Here, as in the subsequent equations in this section, it is understood that each integrand is evaluated at $x = \phi(\X)$.)

Let us now compute the charges associated to the global symmetries discussed in section \ref{sec:globYM} in this extended formalism. We first compute the contraction of the extended presymplectic form with the field space vector $\V_\lambda^{\rm T}$ corresponding to global translations, with $\lambda^\mu$ being the constant vector generating the translation. 
Using the fact that $\If_{\V_{\lambda}^{\rm T}} \chi^\mu = - \lambda^\mu$, and recalling that the contraction of the non-extended presymplectic form $\Omega_{\rm EM}$ has already been calculated in \eqref{iTranslationOmega}, we get
\begin{align}\label{translationcontraction}
	\If_{\V_\lambda^{\rm T}} \Omega^{\rm ext}_{\text{EM}} &= \If_{\V_\lambda^{\rm T}} \Omega_{\text{EM}} - 2\int_{\partial{\cal S}} \dd\sigma_{\mu\nu}\bigl(\phi(\X)\bigr)\, \bigl(\lambda^\mu F^{\nu\rho} \delta A_\rho + \lambda^\sigma \chi^\mu F^{\nu\rho}\partial_\sigma A_\rho - \lambda^\mu \chi^\nu \lem\bigr) \notag \\
	&= \int_{\cal S} \dd\Sigma_\mu\bigl(\phi(\X)\bigr)\, \delta J_T^\mu + 2\int_{\partial{\cal S}} \dd\sigma_{\mu\nu}\bigl(\phi(\X)\bigr)\, \chi^\nu J_T^\mu ,
\end{align}
where we have defined the translation charge current 
\begin{equation}
	J_T^\mu = \lambda^\sigma F^{\mu\nu}\partial_\sigma A_\nu + \lambda^\mu \lem.
\end{equation}
Note that the equations of motion imply that the translation current is conserved:
\begin{equation}
	\partial_\mu J^\mu_T = 0.
	\label{JT-cons}
\end{equation}
Using this fact and equation \eqref{dI} for calculating integral variations, we get
\begin{equation}
	\delta \left(\int_{\cal S} \dd\Sigma_\mu\bigl(\phi(\X)\bigr)\, J^\mu_T\right) = \int_{\cal S} \dd\Sigma_\mu\bigl(\phi(\X)\bigr)\, \delta J_T^\mu + 2\int_{\partial{\cal S}} \dd\sigma_{\mu\nu}\bigl(\phi(\X)\bigr) \chi^\nu J_T^\mu.
	\label{dphi-JT}
\end{equation}
We thus conclude that
\begin{equation}
	\If_{\V^{\rm T}_\lambda} \Omega^{\rm ext}_{\text{EM}} = \delta\left(\int_{\cal S} \dd\Sigma_\mu\bigl(\phi(\X)\bigr)\,\lambda^\sigma  \left(F^{\mu\nu}\partial_\sigma A_\nu + \delta ^\mu_\sigma \lem \right)\right) = \delta \tilde{H}_{\text{EM}}^{\rm T}[\lambda].
\end{equation}
Hence we see that, as we mentioned earlier, the addition of the embedding to the phase space and the use of the extended presymplectic form leads to the elimination of the flux term which appears in \eqref{TranslationFlux}.

Let us then move to the Lorentz symmetries in the extended formalism. Now we have that $\If_{\V^{\text{L}}_\lambda}\chi^\mu = -\lambda\updown{\mu}{\nu}x^\nu$, where $\lambda\updown{\mu}{\nu}$ is the infinitesimal generator of the Lorentz transformation, and the contraction of the extended presymplectic form along a Lorentz transformation orbit gives
\begin{align}
	\If_{\V^{\text{L}}_\lambda} \Omega^{\rm ext}_{\text{EM}} &= \If_{\V^{\text{L}}_\lambda} \Omega_{\text{EM}}-2\int_{\partial{\cal S}} \dd\sigma_{\mu\nu}\bigl(\phi(\X)\bigr) \Bigl(\lambda \updown{\mu}{\rho}x^\rho F^{\nu\alpha} \delta A_\alpha + \chi^\mu F^{\nu\alpha} \lambda \updown{\rho}{\sigma}x^{\sigma} \partial_\rho A_\alpha \notag \\
	&\hspace{150pt} + \chi^\mu F^{\nu\alpha}\lambda \updown{\rho}{\alpha} A_\rho - \chi^\nu \lambda \updown{\mu}{\rho}x^\rho \lem\Bigr) \notag \\
	&= \int_{\cal S} \dd\Sigma_\mu\bigl(\phi(\X)\bigr)\, \delta J^\mu_L - 2 \int_{\partial{\cal S}} \dd\sigma_{\mu\nu}\bigl(\phi(\X)\bigr) \Bigl(\chi^\mu F^{\nu\alpha}\lambda \updown{\rho}{\sigma}x^\sigma\partial_\rho A_\alpha +  \chi^\mu F^{\nu\alpha}\lambda\updown{\rho}{\alpha}A_\rho \notag \\
	&\hspace{200pt} -\lambda \updown{\mu}{\rho}x^\rho \chi^\nu \lem \Bigr) \notag \\
	&= \int_{\cal S} \dd\Sigma_\mu\bigl(\phi(\X)\bigr)\, \delta J^\mu_L + 2\int_{\partial{\cal S}} \dd\sigma_{\mu\nu}\bigl(\phi(\X)\bigr) \chi^\nu J^\mu_L,
\end{align}
where we defined the Lorentz charge current 
\begin{equation}
	J^\mu_L =\lambda\updown{\rho}{\sigma} x^\sigma F^{\mu\nu}\partial_\rho A_\nu +\lambda\updown{\rho}{\sigma} A_\rho F^{\mu\sigma} + \lambda\updown{\mu}{\sigma}x^\sigma \lem.
\end{equation}
A simple calculation shows that this current is also conserved: $\partial_\mu J^\mu_L = 0$. It follows that, as in \eqref{dphi-JT}, we have
\begin{equation}
	\delta \left(\int_{\cal S} \dd\Sigma_\mu\bigl(\phi(\X)\bigr)\,J_L^\mu \right) = \int_{\cal S} \dd\Sigma_\mu\bigl(\phi(\X)\bigr)\, \delta J^\mu_L + 2 \int_{\partial{\cal S}} \dd\sigma_{\mu\nu}\bigl(\phi(\X)\bigr) \chi^\alpha J^\mu_L,
\end{equation}
and we can finally write
\begin{equation}
	\If_{\V_\lambda^{\rm L}} \Omega^{\rm ext}_{\text{EM}} = \delta \left(\int_{\cal S} \dd\Sigma_\mu\bigl(\phi(\X)\bigr)\, \lambda \updown{\rho}{\sigma} \left(x^\sigma F^{\mu\nu}\partial_\rho A_\nu  + \delta^\mu_\rho x^\sigma \lem+A_\rho F^{\mu\sigma}\right)\right) = \delta \tilde{H}_{\text{EM}}^{\rm L}[\lambda].
\end{equation}
As in the translation case, the flux term is eliminated by the extended structure and we only recover the angular-momentum charge of electromagnetism.

To conclude, it is worth noting that the edge mode extension presented in section \ref{sec:extendedYM}, and which guarantees the necessary degeneracy of the presymplectic form in presence of boundaries (see the discussion in section \ref{sec:Lagrangeformalism}), is distinct from the extension developed in this section. The two extensions are compatible with each other and can be treated simultaneously. The two extensions introduce different edge modes, i.e.~fields with contributions associated to the spacetime boundary; the Yang--Mills edge mode in section \ref{sec:extendedYM} is associated to the internal gauge symmetry, while the embedding is associated to spacetime symmetries.

\subsection{Diffeomorphism charges in extended Einstein--Hilbert theory}
\label{sec:boundarycharges}

In the case of the Einstein--Hilbert Lagrangian, the equations of motion imply that $L=0$ on shell. It follows that the expressions \eqref{Theta_ext} and \eqref{Extended_Omega} take a slightly simplified form, namely
\begin{equation}
	\Theta_{\rm EH}^{\rm ext} [g, \delta g, \phi, \delta\phi] = \int_{\cal S} d\Sigma_\mu\bigl(\phi(\X)\bigr)\,\theta_{\rm EH}^\mu\bigl(\phi(\X)\bigr)
\end{equation}
and
\begin{equation}
	\Omega_{\rm EH}^{\rm ext} [g, \delta g, \phi, \delta\phi] = \int_{\cal S} \dd\Sigma_\mu\bigl(\phi(\X)\bigr)\,\omega_{\rm EH}^\mu\bigl(\phi(\X)\bigr) - \int_{\partial{\cal S}} \dd\sigma_{\mu\nu}\bigl(\phi(\X)\bigr)\,\Bigl(\chi^\mu\theta_{\rm EH}^\nu - \chi^\nu\theta_{\rm EH}^\mu\Bigr)\bigl(\phi(\X)\bigr),
\end{equation}
with $\theta^\mu_{\rm EH}$ and $\omega^\mu_{\rm EH}$ given respectively by \eqref{thetaEH} and \eqref{omegaEH1}. In particular, we see that the extended symplectic potential is obtained from its non-extended counterpart by simply acting with the pullback induced by $\phi$, and does not include any additional terms.

Having the extended presymplectic form, we then consider its contraction along gauge directions, i.e.~$\If_\V$, where $\V$ is the tangent vector field of a gauge orbit of diffeomorphisms in the field space and associated to the diffeomorphism generated by a spacetime vector field $\epsilon^\mu$. To carry out the calculations, it is convenient to take $\Omega_{\rm EH}^{\rm ext}$ in the form \eqref{Omega_ext}:
\begin{equation}
	\Omega_{\rm EH}^{\rm ext} = \int_{\cal S} \dd\Sigma_\mu\bigl(\phi(\X)\bigr)\,\Bigl[\omega_{\rm EH}^\mu\bigl(\phi(\X)\bigr) + {\cal L}_\chi\theta_{\rm EH}^\mu\bigl(\phi(\X)\bigr)\Bigr]
	\label{Omega_ext-unsplit}
\end{equation}
where the extended symplectic potential current $\theta^\mu_{\rm ext}$ is equal to $\theta_{\rm EH}^\mu$ for Einstein--Hilbert theory. The contraction of the presymplectic current $\omega_{\rm EH}^\mu$ is calculated in Appendix \ref{sec:gravity_presymplectic_current}. There we established the expression
\begin{equation}
	\If_\V\omega_{\rm EH}^\mu = {\cal L}_\epsilon\theta_{\rm EH}^\mu + \delta\bigl(\nabla_\nu\xi^{\mu\nu}\bigr)
	\label{iomega-simplified}
\end{equation}
where $\xi^{\mu\nu} = \sqrt{-g}\nabla^{[\mu}\epsilon^{\nu]}$; see \eqref{iomega-A2}.
To evaluate the contraction $\If_\V\Omega_{\rm EH}^{\rm ext}$, it now remains to compute the contraction of the Lie derivative
\begin{equation}
	{\cal L}_\chi\theta_{\rm EH}^\mu = \chi^\nu\partial_\nu\theta_{\rm EH}^\mu - (\partial_\nu\chi^\mu)\theta_{\rm EH}^\nu + (\partial_\nu\chi^\nu)\theta_{\rm EH}^\mu.
\end{equation}
Using the fact that $\If_\V\chi^\mu = -\epsilon^\mu$, since $\chi$ is a spacetime vector field, we find (keeping in mind that the contraction $\If_\V$ anticommutes with the field space one-form $\chi^\mu$)
\begin{align}
	\If_\V{\cal L}_\chi\theta_{\rm EH}^\mu &= -\epsilon^\nu\partial_\nu\theta_{\rm EH}^\mu + (\partial_\nu\epsilon^\mu)\theta_{\rm EH}^\nu - (\partial_\nu\epsilon^\nu)\theta_{\rm EH}^\mu \notag \\
	&\quad - \chi^\nu\partial_\nu(\If_\V\theta_{\rm EH}^\mu) + (\partial_\nu\chi^\mu)\If_\V\theta_{\rm EH}^\nu - (\partial_\nu\chi^\nu)\If_\V\theta_{\rm EH}^\mu \notag \\
	&= -{\cal L}_\epsilon\theta_{\rm EH}^\mu - {\cal L}_\chi(\If_\V\theta_{\rm EH}^\mu).
	\label{iLtheta}
\end{align}
Putting together \eqref{iomega-simplified}, \eqref{iLtheta} and \eqref{ithetaEH} for the contraction $\If_\V\theta_{\rm EH}^\mu$, then using \eqref{dI} to identify the integral as a total differential in field space, we arrive at
\begin{align}
	\If_\V\Omega_{\rm EH}^{\rm ext} &= \int_{\cal S} \dd\Sigma_\mu\bigl(\phi(\X)\bigr)\,\Bigl[\delta\bigl(\nabla_\nu\xi^{\mu\nu}\bigr)\bigl(\phi(\X)\bigr) + {\cal L}_\chi\nabla_\nu\xi^{\mu\nu}\bigl(\phi(\X)\bigr)\Bigr] \notag \\
	&= \delta\int_{\cal S} \dd\Sigma_\mu\bigl(\phi(\X)\bigr)\,\nabla_\nu\xi^{\mu\nu}\bigl(\phi(\X)\bigr) \notag \\
	&= \delta\int_{\partial{\cal S}} \dd\sigma_{\mu\nu}\bigl(\phi(\X)\bigr)\,\sqrt{-g\bigl(\phi(\X)\bigr)} \nabla^\mu\epsilon^\nu\bigl(\phi(\X)\bigr).
\end{align}
We therefore obtain
\begin{align}
	\If_\V\Omega_{\rm EH}^{\rm ext} &= \delta H[\epsilon],
	\label{iOmegaExt}
\end{align}
where
\begin{align}
	H[\epsilon]= \int_{\partial{\cal S}} \dd\sigma_{\mu\nu}\bigl(\phi(\X)\bigr)\,\sqrt{-g\bigl(\phi(\X)\bigr)} \nabla^\mu\epsilon^\nu\bigl(\phi(\X)\bigr).
	\label{ExtCharge}
\end{align}
By comparing \eqref{iOmegaExt} to \eqref{iOmegaEH}, we see that the introduction of embedding maps as independent variables in the theory leads to the elimination of the term in \eqref{iOmegaEH} arising from the fact that a generic diffeomorphism does not preserve the boundary. The new boundary charges \eqref{ExtCharge} follow then directly from the contraction of the extended presymplectic form. Aside from the appearance of the embedding maps in the arguments, their expression does not differ from \eqref{ChargeEH} obtained in the standard Einstein--Hilbert theory. The algebra of the new charges is straightforward to calculate and one obtains
\begin{equation}
	\left\{ H[v], H[w] \right\} = \int_{\partial{\cal S}} \dd\sigma_{\mu\nu}\bigl(\phi(\X)\bigr)\,\sqrt{-g\bigl(\phi(\X)\bigr)}\, \nabla^\mu \Bigl(w^\alpha\bigl(\phi(\X)\bigr) \nabla_\alpha v^\nu\bigl(\phi(\X)\bigr) - v^\alpha\bigl(\phi(\X)\bigr) \nabla_\alpha w^\nu\bigl(\phi(\X)\bigr)\Bigr).
	\label{ExtChargesAlg}
\end{equation}

As we will see in the next section, the inclusion of the embedding maps as new variables in the context of gravity allows also to realize the condition that the presymplectic form is degenerate along diffeomorphism orbits, even in the presence of spacetime boundaries. This is the condition we discussed in section \ref{sec:Lagrangeformalism}, and which guarantees the pullback relation between the symplectic form on the gauge reduced field space, and the presymplectic form on the initial field space.

\subsection{Boundary presymplectic form and the degeneracy condition in gravity}
\label{sec:gravity_boundary_symplectic_form}

We will now show that it is possible to manually add a term $\Omega_\phi$ to the presymplectic form, as in the case of Yang--Mills theory in section \ref{sec:extendedYM}, such that the total presymplectic form $\Omega_\text{EH}^\text{T} := \Omega_{\rm EH}^{\rm ext} + \Omega_\phi$ satisfies $\If_\V\Omega_\text{EH}^\text{T} = 0$, where, as before, $\V$ is the tangent vector of a diffeomorphism gauge orbit in field space generated by a spacetime vector field $\epsilon^\mu$. The additional term is defined as an integral over the boundary $\partial{\cal S}$, and it is constructed using the variational vector field $\chi^\mu$ and the embedding map $\phi$. This term arises from the symplectic potential\footnote{This term was first introduced by Donnelly and Freidel in \cite{D.F.2016}.}
\begin{equation}
	\Theta_\phi = -\int_{\partial{\cal S}} \dd\sigma_{\mu\nu}\bigl(\phi(\X)\bigr)\,\sqrt{-g\bigl(\phi(\X)\bigr)}\nabla^\mu\chi^\nu\bigl(\phi(\X)\bigr).
	\label{Theta'}
\end{equation}
Note that
\begin{equation}
	\If_\V\Theta_\phi = \int_{\partial{\cal S}} \dd\sigma_{\mu\nu}\bigl(\phi(\X)\bigr)\,\sqrt{-g\bigl(\phi(\X)\bigr)}\nabla^\mu\epsilon^\nu\bigl(\phi(\X)\bigr) = H[\epsilon].
	\label{ITheta'}
\end{equation}
This implies that the corresponding presymplectic form
\begin{equation}
	\Omega_\phi := \delta\Theta_\phi
	\label{Omega_phi}
\end{equation}
satisfies
\begin{equation}
	\If_\V\Omega_\phi = \Lf_\V\Theta_\phi - \delta \If_\V\Theta_\phi = \Lf_\V\Theta_\phi - \delta H[\epsilon].
\end{equation}
It follows that if the Lie derivative $\Lf_\V\Theta_\phi$ vanishes, we have
\begin{equation}
	\If_\V\Omega_\phi = -\delta H[\epsilon]
\end{equation}
and hence the quantity
\begin{equation}
	\Omega_\text{EH}^\text{T} := \Omega_{\rm EH}^{\rm ext} + \Omega_\phi
	\label{Omega_tot}
\end{equation}
which we define as the total presymplectic form, satisfies
\begin{equation}
	\If_\V\Omega_\text{EH}^\text{T} = 0.
	\label{IOmega_tot=0}
\end{equation}

To establish that \eqref{IOmega_tot=0} holds, we must therefore show that the Lie derivative $\Lf_\V\Theta_\phi$ is equal to zero. (Geometrically this condition indicates the invariance of the symplectic potential $\Theta_\phi$ under spacetime diffeomorphisms.) Via Cartan's formula, the Lie derivative of the symplectic potential \eqref{Theta'} gives
\begin{equation}
	\Lf_\V\Theta_\phi = \delta \If_\V\Theta_\phi + \If_\V\delta\Theta_\phi.
\end{equation}
Applying the basic identity \eqref{dI} to take the variation of \eqref{ITheta'}, we obtain
\begin{equation}
	\delta \If_\V\Theta_\phi = -\int_{\partial{\cal S}} \dd\sigma_{\mu\nu}\bigl(\phi(\X)\bigr)\,\Bigl[\bigl(\delta\xi^{\mu\nu}\bigr)\bigl(\phi(\X)\bigr) + {\cal L}_\chi\xi^{\mu\nu}\bigl(\phi(\X)\bigr)\Bigr]
	\label{dITheta'}
\end{equation}
where the previously introduced abbreviation $\xi^{\mu\nu} = \sqrt{-g}\nabla^{[\mu}\epsilon^{\nu]}$ is used. Similarly, we find
\begin{equation}
	\If_\V\delta\Theta_\phi = \int_{\partial{\cal S}} \dd\sigma_{\mu\nu}\bigl(\phi(\X)\bigr)\,\Bigl[\bigl(\If_\V\delta\eta^{\mu\nu}\bigr)\bigl(\phi(\X)\bigr) + \If_\V\bigl({\cal L}_\chi\eta^{\mu\nu}\bigr)\bigl(\phi(\X)\bigr)\Bigr].
	\label{IdTheta'}
\end{equation}
where we denote
\begin{equation}
	\eta^{\mu\nu} = \sqrt{-g}\nabla^{[\mu}\chi^{\nu]}.
\end{equation}
Now, using $\If_\V\chi^\mu = -\epsilon^\mu$ and $\If_\V\eta^{\mu\nu} = -\xi^{\mu\nu}$, and the explicit expression for the Lie derivative of the tensor density $\eta^{\mu\nu}$, one can see that
\begin{equation}
	\If_\V\bigl({\cal L}_\chi\eta^{\mu\nu}\bigr) = -{\cal L}_\epsilon\eta^{\mu\nu} + {\cal L}_\chi\xi^{\mu\nu}.
\end{equation}
From \eqref{dITheta'} and \eqref{IdTheta'} we then get
\begin{equation}
	\Lf_\V\Theta_\phi = \int_{\partial{\cal S}} \dd\sigma_{\mu\nu}\bigl(\phi(\X)\bigr)\,\Bigl[-\bigl(\delta\xi^{\mu\nu}\bigr)\bigl(\phi(\X)\bigr) + \bigl(\If_\V\delta\eta^{\mu\nu}\bigr)\bigl(\phi(\X)\bigr) - {\cal L}_\epsilon\eta^{\mu\nu}\bigl(\phi(\X)\bigr)\Bigr].
\end{equation}
Further calculations, which due to their length are presented in Appendix \ref{sec:LTheta}, eventually show that the above integrand reduces to an expression which vanishes due to the algebraic symmetries of the Riemann tensor. Hence the conclusion 
\begin{equation}
	\Lf_\V\Theta_\phi = 0
	\label{LTheta'=0}
\end{equation}
is valid as a purely geometrical statement, and it holds independently of whether or not the equations of motion are assumed to be satisfied.

Having \eqref{LTheta'=0}, the argument given in \eqref{Omega_phi}--\eqref{IOmega_tot=0} shows that the presymplectic form $\Omega_\phi := \delta\Theta_\phi$ satisfies
\begin{equation}
	\If_\V\Omega_\phi = -\delta\int_{\partial{\cal S}} \dd\sigma_{\mu\nu}\bigl(\phi(\X)\bigr)\,\sqrt{-g\bigl(\phi(\X)\bigr)}\nabla^\mu\epsilon^\nu\bigl(\phi(\X)\bigr)
\end{equation}
It follows that the total presymplectic form defined in \eqref{Omega_tot} fulfills the condition
\begin{equation}
	\If_\V\Omega_\text{EH}^\text{T} = \If_\V\bigl(\Omega_{\rm EH}^{\rm ext} + \Omega_\phi\bigr) = 0
\end{equation}
showing that $\Omega_{\rm EH}^{\rm T}$ has vanishing components in the directions of diffeomorphism gauge orbits.

\section{Summary and comments}\label{sec:summary}

In the present article, we gave a self-contained review of the covariant phase space formalism and its application to gauge theories with spacetime boundaries. We started by briefly reviewing Cartan calculus on spacetime and its generalization to the field space, employing a simple language and without involving abstract geometrical concepts typically used in the literature. We then exposed some general results in gauge theories, obtained via the covariant phase space formalism, such as Noether's second theorem. We also discussed the degeneracy of the presymplectic form in gauge theories, and we particularly emphasized that the degeneracy along gauge orbits in the field space is a necessary requirement, in order to guarantee that the presymplectic form on the initial field space is the pullback of the symplectic form on the gauge reduced field space. This degeneracy condition was first discussed by Witten and Crnkovi\'c \cite{C.W.1987} in the context of gauge theories without spacetime boundaries. In this article, we extend this requirement to the case where spacetime boundaries are present, and it turns out that in our approach, where we consider only field-independent gauge transformations, this requirement is the source of the emergence of edge modes associated to the boundaries. This is in contrast with the approach in \cite{D.F.2016} for instance, where the edge modes arise as a consequence of requiring gauge invariance of the presymplectic form with respect to field-dependent gauge transformations.

We then moved to applying the formalism to specific gauge theories. We first considered Yang--Mills theory, where we showed that the presymplectic form is inherently invariant with respect to field-independent gauge transformations, even in the presence of spacetime boundaries. However, we also demonstrated that the presymplectic form produces a gauge symmetry charge with support on the boundary. Therefore, instead of imposing gauge invariance, we impose the degeneracy of the presymplectic form along gauge directions. This leads to the introduction of a gauge group-valued field, the Yang--Mills edge mode, which becomes the source of a boundary contribution that cancels the initial boundary charge. We additionally study the global Poincaré symmetry in Yang--Mills theory, exemplified by Maxwell theory. This serves to introduce the notion of fluxes within the familiar context of electrodynamics. We indeed show that these fluxes emerge as a result of the fact that global spacetime symmetries, unlike gauge symmetries, do not preserve the boundary.

In the second example of the application of the covariant phase space formalism to gauge theories, we considered the gravitational theory defined by the Einstein--Hilbert action. We showed that, similarly to the case of global symmetries in Yang--Mills theory, the contraction of the presymplectic form along the diffeomorphism orbits does not give a total differential and fluxes are present. The restriction to diffeomorphisms preserving the boundary provides a charge supported on this boundary and corresponding to the well-known Komar charge. We then illustrate these results by an example where we calculate these charges for the Kerr--Newman--de Sitter spacetime and its limiting cases.

The presence of fluxes, both for global Poincaré symmetry in Yang--Mills theory and diffeomorphism symmetry in the gravitational theory, serves as a motivation to extend the covariant phase space formalism to accommodate symmetries which do not preserve spacetime boundaries. This is accomplished by considering that spacetime regions are defined through embedding maps of some abstract space, and these embedding maps are then included in the theory as new phase space variables. In the case of global symmetries in Yang--Mills theory, we show that the variation of the extended action produces an additional term on the boundary which cancels the fluxes, yielding the established expressions for the canonical energy-momentum and angular-momentum charges.
In the case of gravity, taking the variation of the extended action shows that the extended presymplectic potential is simply the standard one pulled back by the embedding map. However the extended presymplectic form calculated from the potential has an additional co-dimension 2 term. We then demonstrated that the contraction of this extended presymplectic form with any diffeomorphism gives the exact differential of the Komar charge. Furthermore, we show that the additional edge mode term in the case of gravity, required for satisfying the degeneracy condition, can also be expressed in terms of the embedding map. Consequently, we deduce that the edge mode in gravity is the embedding map, and its addition can take care of both the fluxes and the degeneracy condition. The fact that the two extensions coincide in the case of gravity is a consequence of the fact that gravity gauge group is the diffeomorphism group.

In conclusion, we brought forward a different interpretation and justification for the presence of the edge modes in gauge theories with spacetime boundaries. Indeed, in our context where gauge transformations are field-independent, the edge modes appear primarily to realize the degeneracy condition of the presymplectic form along gauge orbits, which in turn allows to connect the symplectic form on the gauge reduced field space with the presymplectic form on the original field space via pullback. Finally, let us note that the dynamics of the edge modes is not yet fully understood. As mentioned in the introduction, in the quantum Hall effect, the additional boundary term is justified by the gauge invariance of the Lagrangian, and this term is added directly to it. As such, the dynamics of the edge mode in that context is well understood. However, in the case of Yang--Mills and gravity theories, the additional boundary term is so far introduced at the level of the presymplectic forms. The Lagrangian of the extended theory which would give rise to such an modification is not yet known. This Lagrangian would provide the dynamics for both the bulk and boundary degrees of freedom (the edge modes), and would also provide a starting point for a path integral formulation of the quantum theory in the presence of boundaries.
This will be the subject of future work.

\subsection*{Acknowledgments}

For IM this work was funded by National Science Centre, Poland grant no. 2022/44/C/ST2/00023. For JKG this work was supported by funds provided by the Polish National Science Center,  the project number  2019/33/B/ST2/00050. For the purpose of open access, the authors have applied a CC BY public copyright license to any author accepted manuscript (AAM) version arising from this submission.

\clearpage

\appendix

\section{Appendix}

\subsection{Contraction of the gravity symplectic current along a gauge orbit} \label{sec:gravity_presymplectic_current}

In this section we perform the calculation of the interior product $\If_\V\omega^\mu_{\rm EH}$, where
\begin{equation}
	\omega_\text{EH}^\mu [g, \delta g] = \frac{1}{2}\left(\delta\bigl(\sqrt{-g}g^{\alpha\beta}\bigr)\delta\Gamma^\mu_{\alpha\beta} - \delta\bigl(\sqrt{-g}g^{\mu\alpha}\bigr)\delta\Gamma^\beta_{\alpha\beta} \right)
	\label{omegaEH-app}
\end{equation}
is the Einstein--Hilbert symplectic current from \eqref{omegaEH1}, and $\V$ is the tangent vector field of a diffeomorphism gauge orbit in field space. We begin by expressing the symplectic current in a form which is more suitable for calculation by carrying out some of the field variations. We have
\begin{equation}
	\delta\bigl(\sqrt{-g}g^{\alpha\beta}\bigr) = \bigl(\delta\sqrt{-g}\bigr)g^{\alpha\beta} + \sqrt{-g}\delta g^{\alpha\beta} = \sqrt{-g}\Bigl(\delta g^{\alpha\beta} + g^{\alpha\beta}\delta\gamma\Bigr).
\end{equation}
where we have introduced the notation
\begin{equation}
	\delta\gamma = \frac{1}{\sqrt{-g}}\delta\sqrt{-g} = \delta\ln\sqrt{-g}
\end{equation}
or, in terms of the metric components,
\begin{equation}
	\delta\gamma = \frac{1}{2}g^{\alpha\beta}\delta g_{\alpha\beta} = -\frac{1}{2}g_{\alpha\beta}\delta g^{\alpha\beta}.
	\label{dgamma}
\end{equation}
Using this in \eqref{omegaEH-app}, we obtain
\begin{equation}
	\omega_\text{EH}^\mu = \frac{1}{2}\sqrt{-g}\Bigl(\delta g^{\alpha\beta}\delta\Gamma^\mu_{\alpha\beta} - \delta g^{\mu\alpha}\delta\Gamma^\beta_{\alpha\beta} + \delta\gamma\bigl(g^{\alpha\beta}\delta\Gamma^\mu_{\alpha\beta} - g^{\mu\alpha}\delta\Gamma^\beta_{\alpha\beta}\bigr)\Bigr).
	\label{omegaEH2}
\end{equation}
In the second and fourth terms in \eqref{omegaEH2} we have, using \eqref{dG},
\begin{equation}
	\delta\Gamma^\beta_{\alpha\beta} = \nabla_\alpha\delta\gamma.
\end{equation}
In the third term there appears the combination
\begin{align}
	g^{\alpha\beta}\delta\Gamma^\mu_{\alpha\beta} &= \frac{1}{2}g^{\alpha\beta}g^{\mu\nu} \Bigl(\nabla_\alpha\delta g_{\nu\beta} + \nabla_\beta\delta g_{\alpha\nu} - \nabla_\nu\delta g_{\alpha\beta}\Bigr) \notag \\
	&= g^{\mu\nu}\nabla^\alpha\delta g_{\alpha\nu} - \frac{1}{2}g^{\alpha\beta}\nabla^\mu\delta g_{\alpha\beta}.
\end{align}
Using the fact that
\begin{equation}
	g^{\mu\nu}\nabla^\alpha\delta g_{\alpha\nu} = -g^{\mu\nu}g_{\alpha\beta}g_{\nu\lambda}\nabla^\alpha\delta g^{\beta\lambda} = -\nabla_\beta\delta g^{\beta\mu}
\end{equation}
we can see that
\begin{equation}
	g^{\alpha\beta}\delta\Gamma^\mu_{\alpha\beta} = -\nabla_\alpha \delta g^{\mu\alpha} - \nabla^\mu\delta\gamma
\end{equation}
and
\begin{equation}
	g^{\alpha\beta}\delta\Gamma^\mu_{\alpha\beta} - g^{\mu\alpha}\delta\Gamma^\beta_{\alpha\beta} = -\nabla_\alpha\delta g^{\mu\alpha} - 2\nabla^\mu\delta\gamma.
\end{equation}
All together, this yields the following expression for the symplectic current:
\begin{align}
	\omega_\text{EH}^\mu [g, \delta g] = \frac{1}{2}\sqrt{-g}\Bigl(\delta g^{\alpha\beta}\delta\Gamma^\mu_{\alpha\beta} - \delta g^{\mu\alpha}\nabla_\alpha\delta\gamma - \delta\gamma\nabla_\alpha\delta g^{\mu\alpha} - 2\delta\gamma\nabla^\mu\delta\gamma\Bigr).
	\label{omegaEH}
\end{align}

Now, in order to compute the interior product of $\omega_\text{EH}^\mu$, we begin by establishing the equations of motion satisfied by the metric variation $\delta g^{\mu\nu}$, as these will be needed in the upcoming calculation. Taking the variation of the Einstein equations
\begin{equation}
	R_{\mu\nu} = 0
\end{equation}
we obtain
\begin{equation}
	\delta R_{\mu\nu} = \nabla_\alpha\delta\Gamma^\alpha_{\mu\nu} - \nabla_\mu\delta\Gamma^\alpha_{\nu\alpha} = 0.
	\label{dR=0}
\end{equation}
Here the first term is
\begin{align}
	\nabla_\alpha\delta\Gamma^\alpha_{\mu\nu} &= \frac{1}{2}g^{\alpha\beta}\nabla_\alpha\Bigl(\nabla_\mu\delta g_{\beta\nu} + \nabla_\nu\delta g_{\mu\beta} - \nabla_\beta\delta g_{\mu\nu}\Bigr) \notag \\
	&= \frac{1}{2}\Bigl(\nabla^\alpha\nabla_\mu\delta g_{\alpha\nu} + \nabla^\alpha\nabla_\nu\delta g_{\alpha\mu} - \nabla^\alpha\nabla_\alpha\delta g_{\mu\nu}\Bigr) \notag \\
	&= -\frac{1}{2}\Bigl(g_{\beta\nu}\nabla_\alpha\nabla_\mu\delta g^{\alpha\beta} + g_{\beta\mu}\nabla_\alpha\nabla_\nu\delta g^{\alpha\beta} - g_{\mu\rho}g_{\nu\sigma}\nabla^\alpha\nabla_\alpha\delta g^{\rho\sigma}\Bigr).
\end{align}
The second term is simply
\begin{equation}
	\nabla_\mu\delta\Gamma^\alpha_{\nu\alpha} = \nabla_\mu\nabla_\nu\delta\gamma.
\end{equation}
Hence we obtain the equations of motion for the metric variation $\delta g^{\mu\nu}$ in the form
\begin{equation}
	g_{\beta\nu}\nabla_\alpha\nabla_\mu\delta g^{\alpha\beta} + g_{\beta\mu}\nabla_\alpha\nabla_\nu\delta g^{\alpha\beta} - g_{\alpha\mu}g_{\beta\nu}\nabla^\lambda\nabla_\lambda\delta g^{\alpha\beta} + 2\nabla_\mu\nabla_\nu\delta\gamma = 0
\end{equation}
or, raising the indices,
\begin{equation}
	\nabla_\alpha\nabla^\mu\delta g^{\alpha\nu} + \nabla_\alpha\nabla^\nu\delta g^{\alpha\mu} - \nabla^\alpha\nabla_\alpha\delta g^{\mu\nu} + 2\nabla^\mu\nabla^\nu\delta\gamma = 0.
	\label{EOM_EH_dg}
\end{equation}
where $\delta \gamma$ is given in terms of the $\delta g^{\mu\nu}$ in \eqref{dgamma}.

Let us also write down the equation obtained by contracting the indices $\mu$ and $\nu$, \ie taking the trace. The contraction yields
\begin{equation}
	2\nabla_\alpha\nabla_\mu\delta g^{\alpha\mu} - g_{\alpha\beta}\nabla^\lambda\nabla_\lambda\delta g^{\alpha\beta} + 2\nabla^\mu\nabla_\mu\delta\gamma = 0.
\end{equation}
and since $\delta\gamma = -\frac{1}{2}g_{\alpha\beta}\delta g^{\alpha\beta}$, the trace of the equations of motion reads
\begin{equation}
	\nabla_\alpha\nabla_\beta\delta g^{\alpha\beta} + 2\nabla^\alpha\nabla_\alpha\delta\gamma = 0.
	\label{TrEOM_GR}
\end{equation}

Having established these preliminary results, let us then turn to the calculation of interior product of the symplectic current $\omega_\text{EH}^\mu$. Applying the operator $\If_\V$ to \eqref{omegaEH}, we obtain
\begin{align}
	\If_\V\omega_\text{EH}^\mu &= \frac{1}{2}\sqrt{-g}\Bigl(\bigl(\If_\V\delta g^{\alpha\beta}\bigr)\delta\Gamma^\mu_{\alpha\beta} - \delta g^{\alpha\beta}\bigl(\If_\V\delta\Gamma^\mu_{\alpha\beta}\bigr) - \bigl(\If_\V\delta g^{\mu\alpha}\bigr)\nabla_\alpha\delta\gamma + \delta g^{\mu\alpha}\nabla_\alpha\bigl(\If_\V\delta\gamma\bigr) \notag \\
	&\quad - \bigl(\If_\V\delta\gamma\bigr)\nabla_\alpha\delta g^{\mu\alpha} + \delta\gamma\nabla_\alpha\bigl(\If_\V\delta g^{\mu\alpha}\bigr) - 2\bigl(\If_\V\delta\gamma\bigr)\nabla^\mu\delta\gamma + 2\delta\gamma\nabla^\mu\bigl(\If_\V\delta\gamma\bigr)\Bigr).
\end{align}
The first term gives
\begin{align}
	\bigl(\If_\V\delta g^{\alpha\beta}\bigr)\delta\Gamma^\mu_{\alpha\beta} &= -\bigl(\nabla^\alpha\epsilon^\beta\bigr)g^{\mu\nu}\Bigl(\nabla_\alpha\delta g_{\nu\beta} + \nabla_\beta\delta g_{\alpha\nu} - \nabla_\nu\delta g_{\alpha\beta}\Bigr) \notag \\
	&= \bigl(\nabla^\alpha\epsilon^\beta\bigr)g^{\mu\nu}\Bigl(g_{\nu\rho}g_{\beta\sigma} \bigl(\nabla_\alpha\delta g^{\rho\sigma}\bigr) + g_{\alpha\rho}g_{\nu\sigma}\bigl(\nabla_\beta\delta g^{\rho\sigma}\bigr) - g_{\alpha\rho}g_{\beta\sigma}\bigl(\nabla_\nu\delta g^{\rho\sigma}\bigr)\Bigr) \notag \\
	&= \bigl(\nabla^\alpha\epsilon_\sigma\bigr)\bigl(\nabla_\alpha\delta g^{\mu\sigma}\bigr) + \bigl(\nabla_\rho\epsilon^\beta\bigr)\bigl(\nabla_\beta\delta g^{\rho\mu}\bigr) - \bigl(\nabla_\rho\epsilon_\sigma\bigr)\bigl(\nabla^\mu\delta g^{\rho\sigma}\bigr) \notag \\
	&= \bigl(\nabla_\alpha\epsilon_\beta\bigr)\Bigl(\nabla^\alpha\delta g^{\mu\beta} + \nabla^\beta\delta g^{\mu\alpha} - \nabla^\mu\delta g^{\alpha\beta}\Bigr).
\end{align}
In the second term we have the contraction
\begin{align}
	\If_\V\delta\Gamma^\mu_{\alpha\beta} &= \frac{1}{2}g^{\mu\nu}\Bigl(\nabla_\alpha\bigl(\If_\V\delta g_{\nu\beta}\bigr) + \nabla_\beta\bigl(\If_\V\delta g_{\alpha\nu}\bigr) - \nabla_\nu\bigl(\If_\V\delta g_{\alpha\beta}\bigr)\Bigr) \notag \\
	&= \frac{1}{2}g^{\mu\nu}\Bigl(\nabla_\alpha\bigl(\nabla_\nu\epsilon_\beta + \nabla_\beta\epsilon_\nu\bigr) + \nabla_\beta\bigl(\nabla_\alpha\epsilon_\nu + \nabla_\nu\epsilon_\alpha\bigr) - \nabla_\nu\bigl(\nabla_\alpha\epsilon_\beta + \nabla_\beta\epsilon_\alpha\bigr)\Bigr) \notag \\
	&= \frac{1}{2}g^{\mu\nu}\Bigl(\bigl[\nabla_\alpha, \nabla_\nu\bigr]\epsilon_\beta + \bigl[\nabla_\beta, \nabla_\nu\bigr]\epsilon_\alpha + \bigl(\nabla_\alpha\nabla_\beta + \nabla_\beta\nabla_\alpha\bigr)\epsilon_\nu \Bigr) \notag \\
	&= \frac{1}{2}\Bigl(\bigl[\nabla_\alpha, \nabla^\mu\bigr]\epsilon_\beta + \bigl[\nabla_\beta, \nabla^\mu\bigr]\epsilon_\alpha + \bigl(\nabla_\alpha\nabla_\beta + \nabla_\beta\nabla_\alpha\bigr)\epsilon^\mu\Bigr) ,
	\label{idGamma}
\end{align}
and therefore
\begin{align}
	\delta g^{\alpha\beta}\bigl(\If_\V\delta\Gamma^\mu_{\alpha\beta}\bigr) &= \delta g^{\alpha\beta}\Bigl(\bigl[\nabla_\alpha, \nabla^\mu\bigr]\epsilon_\beta + \nabla_\alpha\nabla_\beta\epsilon^\mu\Bigr).
\end{align}
Using the fact that
\begin{equation}
	\If_\V\delta\gamma = \frac{1}{2}g^{\alpha\beta}\If_\V\delta g_{\alpha\beta} = \nabla_\alpha\epsilon^\alpha
\end{equation}
we find
\begin{align}
	\If_\V\omega_\text{EH}^\mu &= \frac{1}{2}\sqrt{-g}\Bigl(\bigl(\nabla_\alpha\epsilon_\beta + \nabla_\beta\epsilon_\alpha\bigr)\nabla^\beta\delta g^{\mu\alpha} - \bigl(\nabla_\alpha\epsilon_\beta\bigr)\bigl(\nabla^\mu\delta g^{\alpha\beta}\bigr) \notag \\
		& \hspace{48pt} - \delta g^{\alpha\beta}\Bigl(\bigl[\nabla_\alpha, \nabla^\mu\bigr]\epsilon_\beta - \nabla^\mu\nabla_\alpha\epsilon_\beta + \nabla_\alpha\nabla_\beta\epsilon^\mu\Bigr) \notag \\
		&\hspace{48pt} - \bigl(-\nabla^\mu\epsilon^\alpha - \nabla^\alpha\epsilon^\mu\bigr)\bigl(\nabla_\alpha\delta\gamma\bigr) + \delta g^{\mu\alpha}\nabla_\alpha\bigl(\nabla_\beta\epsilon^\beta\bigr) - \bigl(\nabla_\beta\epsilon^\beta\bigr)\bigl(\nabla_\alpha\delta g^{\mu\alpha}\bigr) \notag \\
	&\hspace{48pt}  + \delta\gamma\nabla_\alpha\bigl(-\nabla^\mu\epsilon^\alpha - \nabla^\alpha\epsilon^\mu\bigr) - 2\bigl(\nabla_\alpha\epsilon^\alpha\bigr)\bigl(\nabla^\mu\delta\gamma\bigr) + 2\delta\gamma\nabla^\mu\bigl(\nabla_\alpha\epsilon^\alpha\bigr)\Bigr).
	\label{iomegaEH2}
\end{align}
In the following, we simplify the expression of $\If_\V \omega_\text{EH}^\mu$ in \eqref{iomegaEH2} via several manipulations involving the use of the equations of motion. 

Collecting similar terms together based on the position of the free index $\mu$, and noting that the equation of motion $R_{\mu\nu} = 0$ implies
\begin{equation}
	\nabla^\mu\nabla_\alpha\epsilon^\alpha - \nabla_\alpha\nabla^\mu\epsilon^\alpha = g^{\mu\nu}R\updown{\alpha}{\beta\nu\alpha}\epsilon^\beta = -R\downup{\beta}{\mu}\epsilon^\beta = 0,
\end{equation}
we arrange the expression \eqref{iomegaEH2} as
\begin{align}
	\If_\V\omega_\text{EH}^\mu = \frac{1}{2}\sqrt{-g}\biggl[&-\bigl(\nabla_\alpha\nabla_\beta\epsilon^\mu\bigr)\delta g^{\alpha\beta} - \bigl(\nabla^\alpha\nabla_\alpha\epsilon^\mu\bigr)\delta\gamma + \bigl(\nabla^\alpha\epsilon^\mu\bigr)\nabla_\alpha\delta\gamma \notag \\
		& + \bigl(\nabla_\alpha\epsilon_\beta + \nabla_\beta\epsilon_\alpha\bigr)\nabla^\beta\delta g^{\mu\alpha} - \bigl(\nabla_\beta\epsilon^\beta\bigr)\nabla_\alpha\delta g^{\mu\alpha} + \bigl(\nabla_\alpha\nabla_\beta\epsilon^\beta\bigr)\delta g^{\mu\alpha} \notag \\[1ex]
		& - \bigl(\nabla_\alpha\epsilon_\beta\bigr)\nabla^\mu\delta g^{\alpha\beta} + \Bigl(\bigl[\nabla^\mu, \nabla_\alpha\bigr]\epsilon_\beta\Bigr) \delta g^{\alpha\beta}  + \bigl(\nabla^\mu\epsilon^\alpha\bigr)\nabla_\alpha\delta\gamma \notag \\
	& - 2\bigl(\nabla_\alpha\epsilon^\alpha\bigr)\nabla^\mu\delta\gamma + \bigl(\nabla^\mu\nabla_\alpha\epsilon^\alpha\bigr)\delta\gamma\biggr].
	\label{iomegaEH3}
\end{align}
Following \cite{C.W.1987}, but paying careful attention to signs, we can establish that on-shell we obtain
\begin{align}
	\If_\V\omega_\text{EH}^\mu = \frac{1}{2}\sqrt{-g} \nabla_\alpha \biggl[ & \epsilon^\mu\bigl(\nabla_\beta\delta g^{\alpha\beta} + 2\nabla^\alpha\delta\gamma\bigr) + \epsilon_\beta\nabla^\alpha\delta g^{\mu\beta} - \bigl(\nabla_\beta\epsilon^\mu\bigr)\delta g^{\alpha\beta} - \bigl(\nabla^\alpha\epsilon^\mu\bigr)\delta\gamma \notag \\
	& - \epsilon^\alpha\bigl(\nabla_\beta\delta g^{\mu\beta} + 2\nabla^\mu\delta\gamma\bigr) - \epsilon_\beta\nabla^\mu\delta g^{\alpha\beta} + \bigl(\nabla_\beta\epsilon^\alpha\bigr)\delta g^{\mu\beta} + \bigl(\nabla^\mu\epsilon^\alpha\bigr)\delta\gamma \biggr] ,
	\label{iomegaEH4}
\end{align} 
showing that $\If_\V\omega_\text{EH}^\mu$ can be expressed as a total divergence of an antisymmetric tensor density. 

To verify this result, we expand the total derivative in the above expression and see that we recover the right-hand side of \eqref{iomegaEH3}, assuming the equations of motion are satisfied. Consider first the terms in \eqref{iomegaEH4} in which the free index is on $\epsilon^\mu$. Taking into account \eqref{TrEOM_GR}, we see that these terms give
\begin{align}
	&\epsilon^\mu\nabla_\alpha\Bigl(\nabla_\beta\delta g^{\alpha\beta} + 2\nabla^\alpha\delta\gamma\Bigr) + \bigl(\nabla_\alpha\epsilon^\mu\bigr)\Bigl(\nabla_\beta\delta g^{\alpha\beta} + 2\nabla^\alpha\delta\gamma\Bigr) \notag \\
	& - \bigl(\nabla_\alpha\nabla_\beta\epsilon^\mu\bigr)\delta g^{\alpha\beta} - \bigl(\nabla_\beta\epsilon^\mu\bigr)\nabla_\alpha\delta g^{\alpha\beta} - \bigl(\nabla_\alpha\nabla^\alpha\epsilon^\mu\bigr)\delta\gamma - \bigl(\nabla^\alpha\epsilon^\mu\bigr)\nabla_\alpha\delta\gamma \notag \\[2ex]
	& = \bigl(\nabla_\alpha\epsilon^\mu\bigr)\nabla^\alpha\delta\gamma -\bigl(\nabla_\alpha\nabla_\beta\epsilon^\mu\bigr)\delta g^{\alpha\beta} - \bigl(\nabla_\alpha\nabla^\alpha\epsilon^\mu\bigr)\delta\gamma ,
\end{align}
which agrees with the terms on the first line of \eqref{iomegaEH3}. 

We then take the terms in \eqref{iomegaEH4} where the free index is on $\delta g^{\mu\beta}$:
\begin{align}
	&\epsilon^\alpha\bigl[\nabla_\alpha, \nabla_\beta\bigr]\delta g^{\mu\beta} + \bigl(\nabla_\alpha\epsilon_\beta\bigr)\nabla^\alpha\delta g^{\mu\beta} + \epsilon_\beta\nabla_\alpha\nabla^\alpha\delta g^{\mu\beta}  \notag \\
	&- \bigl(\nabla_\alpha\epsilon^\alpha\bigr)\nabla_\beta\delta g^{\mu\beta} - \epsilon^\alpha\nabla_\alpha\nabla_\beta\delta g^{\mu\beta} + \bigl(\nabla_\alpha\nabla_\beta\epsilon^\alpha\bigr)\delta g^{\mu\beta} + \bigl(\nabla_\beta\epsilon^\alpha\bigr)\nabla_\alpha\delta g^{\mu\beta} \notag \\[2ex]
	&= -\epsilon^\alpha\nabla_\beta\nabla_\alpha\delta g^{\mu\beta} + \epsilon_\beta\nabla_\alpha\nabla^\alpha\delta g^{\mu\beta} \notag \\
	&\quad + \bigl(\nabla_\alpha\epsilon_\beta + \nabla_\beta\epsilon_\alpha\bigr)\nabla^\alpha\delta g^{\mu\beta} - \bigl(\nabla_\alpha\epsilon^\alpha\bigr)\nabla_\beta\delta g^{\mu\beta} + \bigl(\nabla_\alpha\nabla_\beta\epsilon^\alpha\bigr)\delta g^{\mu\beta}.
	\label{iomega_dg}
\end{align}
Due to the on-shell relation $\nabla_\alpha\nabla_\beta\epsilon^\alpha = \nabla_\beta\nabla_\alpha\epsilon^\alpha$, we see that the terms on the second line reproduce the second line of \eqref{iomegaEH3}. Using the equations of motion \eqref{EOM_EH_dg}, the terms on the first line of \eqref{iomega_dg} become
\begin{align}
	\epsilon_\beta\Bigl(-\nabla_\alpha\nabla^\beta\delta g^{\mu\alpha} + \nabla_\alpha\nabla^\alpha\delta g^{\mu\beta}\Bigr) = \epsilon_\beta\Bigl(\nabla_\alpha\nabla^\mu\delta g^{\alpha\beta} + 2\nabla^\beta\nabla^\mu\delta\gamma\Bigr).
	\label{second}
\end{align}
In \eqref{iomegaEH4} we are now left with the terms where the free index is on a covariant derivative:
\begin{align}
	&\Bigl(\bigl[\nabla^\mu, \nabla_\alpha\bigr]\epsilon_\beta\Bigr)\delta g^{\alpha\beta} - 2\bigl(\nabla_\alpha\epsilon^\alpha\bigr)\nabla^\mu\delta\gamma - 2\epsilon^\alpha\nabla_\alpha\nabla^\mu\delta\gamma \notag \\
	&- \bigl(\nabla_\alpha\epsilon_\beta\bigr)\nabla^\mu\delta g^{\alpha\beta} - \epsilon_\beta\nabla_\alpha\nabla^\mu\delta g^{\alpha\beta} + \bigl(\nabla_\alpha\nabla^\mu\epsilon^\alpha\bigr)\delta\gamma + \bigl(\nabla^\mu\epsilon^\alpha\bigr)\nabla_\alpha\delta\gamma \notag \\[2ex]
	&= \epsilon_\beta\Bigl(-\nabla_\alpha\nabla^\mu\delta g^{\alpha\beta} - 2\nabla^\beta\nabla^\mu\delta\gamma\Bigr) + \Bigl(\bigl[\nabla^\mu, \nabla_\alpha\bigr]\epsilon_\beta\Bigr)\delta g^{\alpha\beta} \notag \\
	&\quad - 2\bigl(\nabla_\alpha\epsilon^\alpha\bigr)\nabla^\mu\delta\gamma - \bigl(\nabla_\alpha\epsilon_\beta\bigr)\nabla^\mu\delta g^{\alpha\beta} + \bigl(\nabla_\alpha\nabla^\mu\epsilon^\alpha\bigr)\delta\gamma + \bigl(\nabla^\mu\epsilon^\alpha\bigr)\nabla_\alpha\delta\gamma .
\end{align}
Here the first two terms cancel the terms left over from the previous step. The remaining terms coincide with the last two lines of \eqref{iomegaEH3} (keeping again in mind that $\nabla_\alpha\nabla^\mu\epsilon^\alpha = \nabla^\mu\nabla_\alpha\epsilon^\alpha$ on-shell). 

Finally, we are left with the terms
\begin{equation}
	\Bigl(\bigl[\nabla^\mu, \nabla_\alpha\bigr]\epsilon_\beta\Bigr)\delta g^{\alpha\beta} + \epsilon^\alpha\bigl[\nabla_\alpha, \nabla_\beta\bigr]\delta g^{\mu\beta} .
	\label{bulk}
\end{equation} 
Using the identity $[\nabla_\alpha, \nabla_\beta]v^\mu = R\updown{\mu}{\nu\alpha\beta}v^\nu$ and its generalization for tensors of rank two, one can easily verify that these terms vanish on-shell, hence establishing that the expression \eqref{iomegaEH4} is correct. 

The result \eqref{iomegaEH4} can be written in a more compact form if we recognize that the terms on the right-hand side are related to the symplectic potential current $\theta^\mu_{\rm EH}$ and the tensor density $\xi^{\mu\nu} = \sqrt{-g}\nabla^{[\mu}\epsilon^{\nu]}$ introduced in \eqref{xi}. The symplectic potential current, which was defined in \eqref{thetaEH}, is equivalently given by the expression
\begin{equation}
	\theta^\mu_{\rm EH} = -\frac{1}{2}\sqrt{-g}\Bigl(\nabla_\alpha\delta g^{\mu\alpha} + 2\nabla^\mu\delta\gamma\Bigr).
	\label{thetaEH-alt}
\end{equation}
Furthermore, direct calculation of the variation $\delta\xi^{\mu\nu}$ shows that
\begin{align}
	\delta\xi^{\mu\nu} &= \sqrt{-g}\Bigl(\nabla^{[\mu}\epsilon^{\nu]}\,\delta\gamma + \delta g^{\alpha[\mu}\nabla_\alpha\epsilon^{\nu]} + g^{\alpha[\mu}\delta\Gamma^{\nu]}_{\alpha\beta}\epsilon^\beta\Bigr) \notag \\
	&= \sqrt{-g}\Bigl(\nabla^{[\mu}\epsilon^{\nu]}\delta\gamma - \nabla_\alpha\epsilon^{[\mu}\delta g^{\nu]\alpha} - \epsilon_\alpha\nabla^{[\mu}\delta g^{\nu]\alpha}\Bigr)
	\label{dxi}
\end{align}
where \eqref{dG} was used to obtain the expression on the second line. Comparing now \eqref{thetaEH-alt} and \eqref{dxi} with \eqref{iomegaEH4}, we see that the contraction $\If_\V\omega^\mu_{\rm EH}$ can be expressed in the simple form
\begin{equation}
	\If_\V\omega_\text{EH}^\mu = \partial_\nu\Bigl(-\epsilon^\mu\theta_\text{EH}^\nu + \epsilon^\nu\theta_\text{EH}^\mu + \delta\xi^{\mu\nu}\Bigr).
	\label{iomega-A1}
\end{equation}

A yet different expression for $\If_\V\omega^\mu_{\rm EH}$, which will be needed in section \ref{sec:boundarycharges}, can be obtained as follows. Observe that the Lie derivative of the vector density $\theta^\mu_{\rm EH}$ can be written on-shell as
\begin{equation}
	{\cal L}_\epsilon\theta^\mu_{\rm EH} = \epsilon^\nu\partial_\nu\theta_{\rm EH}^\mu - (\partial_\nu\epsilon^\mu)\theta_{\rm EH}^\nu + (\partial_\nu\epsilon^\nu)\theta_{\rm EH}^\mu = -\partial_\nu\bigl(\epsilon^\mu\theta_{\rm EH}^\nu - \epsilon^\nu\theta_{\rm EH}^\mu\bigr),
\end{equation}
since the symplectic potential current satisfies $\partial_\mu\theta^\mu_{\rm EH} = 0$ on-shell. Observe also that for the antisymmetric tensor density $\xi^{\mu\nu}$, the covariant divergence coincides with the partial divergence:
\begin{align}
	\nabla_\nu \xi^{\mu\nu} = \partial_\nu \xi^{\mu\nu} + \Gamma^\mu_{\nu\beta}\xi^{\beta\nu} + \Gamma^\nu_{\nu\beta}\xi^{\mu\beta} - \Gamma^\beta_{\nu\beta}\xi^{\mu\nu} = \partial_\nu\xi^{\mu\nu}.
\end{align}
This implies
\begin{equation}
	\partial_\nu\delta\xi^{\mu\nu} = \delta\bigl(\partial_\nu\xi^{\mu\nu}\bigr) = \delta\bigl(\nabla_\nu\xi^{\mu\nu}\bigr),
\end{equation}
and so we have established the alternative expression
\begin{equation}
	\If_\V\omega_{\rm EH}^\mu = {\cal L}_\epsilon\theta_{\rm EH}^\mu + \delta\bigl(\nabla_\nu\xi^{\mu\nu}\bigr)
	\label{iomega-A2}
\end{equation}
for the contraction of the symplectic current.

\subsection{Variation of an embedded integral}
\label{sec:embedded_variation}

Consider the integral
\begin{equation}
	{\cal I}[\alpha] = \int_S \alpha
	\label{I[a]}
\end{equation}
where
\begin{equation}
	\alpha(x) = \alpha_{\mu_1\cdots\mu_p}(x)\,\dd x^{\mu_1}\wedge\cdots\wedge \dd x^{\mu_p}
\end{equation}
is a $p$-form, and $S$ is a $p$-dimensional submanifold of the $d$-dimensional spacetime manifold $M$. Using the Levi-Civita symbol, we can identify the $p$-form $\alpha$ with a tensor density of rank $n := d - p$ and density weight $+1$. Hence the integral \eqref{I[a]} can be written as
\begin{equation}
	{\cal I}[\alpha] = \int_S \dd\Sigma_{\mu_1\cdots\mu_n}(x)\,\alpha^{\mu_1\cdots\mu_n}(x)
\end{equation}
where $\alpha^{\mu_1\cdots\mu_n} = \frac{1}{p!}\epsilon^{\mu_1\cdots\mu_n\mu_{n+1}\cdots\mu_d}\alpha_{\mu_{n+1}\cdots\mu_d}$, and
\begin{equation}
	d\Sigma_{\mu_1\cdots\mu_n}(x) = \epsilon_{\mu_1\cdots\mu_n\mu_{n+1}\cdots\mu_d}\dd x^{\mu_{n+1}}\wedge\cdots\wedge \dd x^{\mu_d}.
\end{equation}
Let $\phi: {\cal D}\subset \R^d \to M$ be an embedding map. We denote $\phi({\cal S}) = S$, \ie ${\cal S}$ is the preimage of the submanifold $S$ under $\phi$. Given such an embedding map, we can construct the pullback of the integral \eqref{I[a]}, thus obtaining
\begin{equation}
	{\cal I}[\alpha, \phi] = \int_{\cal S} \phi^*\alpha = \int_{\cal S} \dd\Sigma_{\mu_1\cdots\mu_n}\bigl(\phi(\X)\bigr)\,\alpha^{\mu_1\cdots\mu_n}\bigl(\phi(\X)\bigr).
	\label{I[a,f]}
\end{equation}
(Here $\X$ denotes a set of coordinates on $\D$.) 

In this section we wish to establish a formula for the variation -- \ie field space differential -- of an integral of this form, assuming that both the $p$-form $\alpha$ and the embedding map $\phi$ are field space dependent variables. 

Let us begin by explicitly extracting the $\phi$-dependence out of the surface element $\dd\Sigma_{\mu_1\cdots\mu_n}\bigl(\phi(\X)\bigr)$. We have
\begin{align}
	\dd\Sigma_{\mu_1\cdots\mu_n}\bigl(\phi(\X)\bigr) &= \epsilon_{\mu_1\cdots\mu_n\mu_{n+1}\cdots\mu_d}\,\dd\phi^{\mu_{n+1}}(\X)\wedge\cdots\wedge \dd\phi^{\mu_d}(\X) \notag \\
	&= \epsilon_{\mu_1\cdots\mu_n\mu_{n+1}\cdots\mu_d}\,\frac{\partial\phi^{\mu_{n+1}}(\X)}{\partial \X^{\nu_{n+1}}}\cdots\frac{\partial\phi^{\mu_d}(\X)}{\partial \X^{\nu_d}}\dd\X^{\nu_{n+1}}\wedge\cdots\wedge \dd\X^{\nu_d} \notag \\
	&= J(\X)\frac{\partial \X^{\nu_1}}{\partial\phi^{\mu_1}(\X)}\cdots\frac{\partial \X^{\nu_n}}{\partial\phi^{\mu_n}(\X)}\,\dd\Sigma_{\nu_1\cdots\nu_n}(\X)
\end{align}
where
\begin{equation}
	J(\X) = \det\biggl(\frac{\partial\phi^\mu(\X)}{\partial \X^\nu}\biggr)
\end{equation}
is the Jacobian. Denoting the product of partial derivatives as
\begin{equation}
	K\updown{\mu_1\cdots\mu_n}{\nu_1\cdots\nu_n}(\X) = \frac{\partial \X^{\mu_1}}{\partial\phi^{\nu_1}(\X)}\cdots\frac{\partial \X^{\mu_n}}{\partial\phi^{\nu_n}(\X)}
\end{equation}
the integral \eqref{I[a,f]} takes the form
\begin{equation}
	{\cal I}[\alpha, \phi] = \int_{\cal S} \dd\Sigma_{\mu_1\cdots\mu_n}(\X)\,J(\X)K\updown{\mu_1\cdots\mu_n}{\nu_1\cdots\nu_n}(\X)\alpha^{\nu_1\cdots\nu_n}\bigl(\phi(\X)\bigr).
\end{equation}
Applying the field space exterior derivative $\delta$, we now obtain
\begin{align}
	\delta{\cal I}[\alpha, \phi] &= \int_{\cal S} \dd\Sigma_{\mu_1\cdots\mu_n}(\X)\,\biggl[\delta J(\X)K\updown{\mu_1\cdots\mu_n}{\nu_1\cdots\nu_n}(\X)\alpha^{\nu_1\cdots\nu_n}\bigl(\phi(\X)\bigr) \notag \\
	&+ J(\X)\Bigl(\delta K\updown{\mu_1\cdots\mu_n}{\nu_1\cdots\nu_n}(\X) \alpha^{\nu_1\cdots\nu_n}\bigl(\phi(\X)\bigr) + K\updown{\mu_1\cdots\mu_n}{\nu_1\cdots\nu_n}(\X) \delta\bigl[\alpha^{\nu_1\cdots\nu_n}\bigl(\phi(\X)\bigr)\bigr]\Bigr)\biggr].
	\label{dI[a,f]}
\end{align}
Consider first the variation of $\alpha\bigl(\phi(\X)\bigr)$:
\begin{equation}
	\delta\bigl[\alpha^{\nu_1\cdots\nu_n}\bigl(\phi(\X)\bigr)\bigr] = (\delta\alpha)^{\nu_1\cdots\nu_n}\bigl(\phi(\X)\bigr) + \delta\phi^\mu(\X)\partial_\mu\alpha^{\nu_1\cdots\nu_n}\bigl(\phi(\X)\bigr).
	\label{da(f)}
\end{equation}
(Throughout this section, $\partial_\mu$ denotes differentiation with respect to $\phi^\mu(\X) = x^\mu$.) Introducing the object
\begin{equation}
	\chi^\mu(x) := \bigl(\delta\phi^\mu \circ \phi^{-1}\bigr)(x),
\end{equation}
which is a vector field on spacetime and a one-form on field space, the variation \eqref{da(f)} can be written as
\begin{equation}
	\delta\bigl[\alpha^{\nu_1\cdots\nu_n}\bigl(\phi(\X)\bigr)\bigr] = (\delta\alpha)^{\nu_1\cdots\nu_n}\bigl(\phi(\X)\bigr) + \chi^\mu\bigl(\phi(\X)\bigr)\partial_\mu\alpha^{\nu_1\cdots\nu_n}\bigl(\phi(\X)\bigr).
\end{equation}
Next, we deduce the variation of the partial derivatives by using the matrix identity
\begin{equation}
	\delta A^{-1} = -A^{-1}(\delta A)A^{-1}.
\end{equation}
We have
\begin{align}
	\delta\biggl(\frac{\partial \X^\mu}{\partial \phi^\nu(\X)}\biggr) &= -\frac{\partial \X^\mu}{\partial \phi^\lambda(\X)}\frac{\partial\delta \phi^\lambda(\X)}{\partial \X^\sigma}\frac{\partial \X^\sigma}{\partial \phi^\nu(\X)} \notag \\
	&= -\frac{\partial \X^\mu}{\partial \phi^\lambda(\X)}\frac{\partial\chi^\lambda\bigl(\phi(\X)\bigr)}{\partial \X^\sigma}\frac{\partial \X^\sigma}{\partial \phi^\nu(\X)} \notag \\
	&= -\frac{\partial \X^\mu}{\partial \phi^\lambda(\X)}\partial_\nu\chi^\lambda\bigl(\phi(\X)\bigr)
\end{align}
from which it follows that
\begin{align}
	\delta K\updown{\mu_1\cdots\mu_n}{\nu_1\cdots\nu_n}(\X) =\; &{-} \partial_{\nu_1}\chi^\lambda\bigl(\phi(\X)\bigr)K\updown{\mu_1\cdots\mu_n}{\lambda\nu_2\cdots\nu_n}(\X) - \ldots \notag \\
	&{-}\partial_{\nu_n}\chi^\lambda\bigl(\phi(\X)\bigr) K\updown{\mu_1\cdots\mu_n}{\nu_1\cdots\nu_{n-1}\lambda}(\X).
\end{align}
Finally, the variation of the Jacobian is obtained by applying the identity
\begin{equation}
	\delta\det A = (\det A)\Tr\bigl(A^{-1}\delta A\bigr)
\end{equation}
to the matrix
\begin{equation}
	A\updown{\mu}{\nu} = \frac{\partial \phi^\mu(\X)}{\partial \X^\nu}.
\end{equation}
Since
\begin{equation}
	\Tr\bigl(A^{-1}\delta A\bigr) = (A^{-1})\updown{\mu}{\nu}\delta A\updown{\nu}{\mu} = \frac{\partial \X^\mu}{\partial \phi^\nu(\X)}\frac{\partial\delta \phi^\nu(\X)}{\partial \X^\mu} = \frac{\partial\delta \phi^\nu(\X)}{\partial \phi^\nu(\X)} = \partial_\nu\chi^\nu\bigl(\phi(\X)\bigr) ,
\end{equation}
we see that
\begin{equation}
	\delta J(\X) = J(\X)\partial_\mu\chi^\mu\bigl(\phi(\X)\bigr).
\end{equation}
Now going back to \eqref{dI[a,f]}, we find
\begin{align}
	\delta{\cal I}[\alpha, \phi] = \int_{\cal S} &d\Sigma_{\mu_1\cdots\mu_n}(\X)\,J(\X)K\updown{\mu_1\cdots\mu_n}{\nu_1\cdots\nu_n}(\X)\biggl[(\delta\alpha)^{\nu_1\cdots\nu_n}\bigl(\phi(\X)\bigr) \notag \\
		&+ \partial_\lambda\chi^\lambda\bigl(\phi(\X)\bigr) \alpha^{\nu_1\cdots\nu_n}\bigl(\phi(\X)\bigr) + \chi^\lambda\bigl(\phi(\X)\bigr) \partial_\lambda\alpha^{\nu_1\cdots\nu_n}\bigl(\phi(\X)\bigr) \notag \\
	&- \partial_\lambda\chi^{\nu_1}\bigl(\phi(\X)\bigr) \alpha^{\lambda\nu_2\cdots\nu_n}\bigl(\phi(\X)\bigr) - \ldots - \partial_\lambda\chi^{\nu_n}\bigl(\phi(\X)\bigr)\alpha^{\nu_1\cdots\nu_{n-1}\lambda}\bigl(\phi(\X)\bigr) \biggr] .
\end{align}
This result can be expressed in a more elegant form by observing that the terms involving the vector field $\chi^\mu$ amount to the Lie derivative of the tensor density $\alpha^{\mu_1\cdots\mu_n}$ along $\chi^\mu$:
\begin{equation}
	{\cal L}_\chi\alpha^{\mu_1\cdots\mu_n} = \chi^\lambda\partial_\lambda\alpha^{\mu_1\cdots\mu_n} - (\partial_\lambda\chi^{\mu_1})\alpha^{\lambda\mu_2\cdots\mu_n} - \ldots - (\partial_\lambda\chi^{\mu_n})\alpha^{\mu_1\cdots\mu_{n-1}\lambda} + (\partial_\lambda\chi^\lambda)\alpha^{\mu_1\cdots\mu_n}.
\end{equation}
Hence we have established that the variation of the integral \eqref{I[a,f]} is given by the expression
\begin{equation}
	\delta{\cal I}[\alpha, \phi] = \int_{\cal S} \dd\Sigma_{\mu_1\cdots\mu_n}\bigl(\phi(\X)\bigr)\,\Bigl[(\delta\alpha)^{\mu_1\cdots\mu_n}\bigl(\phi(\X)\bigr) + {\cal L}_\chi\alpha^{\mu_1\cdots\mu_n}\bigl(\phi(\X)\bigr)\Bigr]
\end{equation}
or, in coordinate-free notation,
\begin{equation}
	\delta\int_{\cal S} \phi^*\alpha = \int_{\cal S} \phi^*\bigl(\delta\alpha + {\cal L}_\chi\alpha\bigr).
\end{equation}
Note that the role of the variational vector field $\chi^\mu$ is essentially to encode the contribution arising from the variation of the embedding map; if the map $\phi$ carries no field space dependence we have $\chi^\mu = 0$ and in this case the variation commutes with the pullback as one would expect.

\subsection{Lie derivative of the boundary symplectic potential in gravity}
\label{sec:LTheta}

In this section we present the details of the calculation which shows that the boundary presymplectic term introduced in section \ref{sec:gravity_boundary_symplectic_form} has a vanishing field space Lie derivative along the gauge orbits of diffeomorphisms. In section \ref{sec:gravity_boundary_symplectic_form} we arrived at the expression
\begin{equation}
	\Lf_\V\Theta_\phi = \int_{\partial\Sigma} \dd\sigma_{\mu\nu}\bigl(\phi(\X)\bigr)\,\Bigl[-\bigl(\delta\xi^{\mu\nu}\bigr)\bigl(\phi(\X)\bigr) + \bigl(\If_\V\delta\eta^{\mu\nu}\bigr)\bigl(\phi(\X)\bigr) - {\cal L}_\epsilon\eta^{\mu\nu}\bigl(\phi(\X)\bigr)\Bigr]
	\label{LQ'}
\end{equation}
where
\begin{align}
	\xi^{\mu\nu} &= \sqrt{-g}\nabla^\mu\epsilon^\nu, \\[1ex]
	\eta^{\mu\nu} &= \sqrt{-g}\nabla^\mu\chi^\nu.
\end{align}
(For convenience we omit the antisymmetrization on $\mu$ and $\nu$; this can be done freely, since the expressions we handle will eventually be contracted against the antisymmetric surface element $\dd\sigma_{\mu\nu}$.) For the variation of $\xi^{\mu\nu}$, we have
\begin{equation}
	\delta\xi^{\mu\nu} = \delta\bigl(\sqrt{-g}\nabla^\mu\epsilon^\nu\bigr) = \sqrt{-g}\Bigl(\delta\gamma\nabla^\mu\epsilon^\nu + \delta\bigl(\nabla^\mu\epsilon^\nu\bigr)\Bigr).
\end{equation}
The variation of the covariant derivative gives
\begin{equation}
	\delta\bigl(\nabla^\mu\epsilon^\nu\bigr) = \delta\Bigl(g^{\mu\alpha}\bigl(\partial_\alpha\epsilon^\nu + \Gamma^\nu_{\alpha\beta}\epsilon^\beta\bigr)\Bigr) = \delta g^{\mu\alpha}\nabla_\alpha\epsilon^\nu + g^{\mu\alpha}\delta\Gamma^\nu_{\alpha\beta}\epsilon^\beta
\end{equation}
so we obtain
\begin{equation}
	\delta\xi^{\mu\nu} = \sqrt{-g}\Bigl(\bigl(\nabla^\mu\epsilon^\nu\bigr)\delta\gamma + \bigl(\nabla_\alpha\epsilon^\nu\bigr)\delta g^{\mu\alpha} + g^{\mu\alpha}\epsilon^\beta\delta\Gamma^\nu_{\alpha\beta}\Bigr).
	\label{deta}
\end{equation}
Next, we compute
\begin{equation}
	\delta\eta^{\mu\nu} = \sqrt{-g}\Bigl(\delta\gamma\nabla^\mu\chi^\nu + \delta\bigl(\nabla^\mu\chi^\nu\bigr)\Bigr)
\end{equation}
and
\begin{align}
	\If_\V\delta\eta^{\mu\nu} &= \sqrt{-g}\Bigl(\bigl(\If_\V\delta\gamma\bigr)\nabla^\mu\chi^\nu - \delta\gamma\nabla^\mu\bigl(\If_\V\chi^\nu\bigr) + \If_\V\delta\bigl(\nabla^\mu\chi^\nu\bigr)\Bigr) \notag \\
	&= \sqrt{-g}\Bigl(\bigl(\nabla_\alpha\epsilon^\alpha\bigr)\bigl(\nabla^\mu\chi^\nu\bigr) + \bigl(\nabla^\mu\epsilon^\nu\bigr)\delta\gamma + \If_\V\delta\bigl(\nabla^\mu\chi^\nu\bigr)\Bigr).
	\label{Idxi}
\end{align}
In the variation of the covariant derivative, the field space exterior derivative now acts also on the variational vector field $\chi^\mu$:
\begin{align}
	\delta\bigl(\nabla^\mu\chi^\nu\bigr) &= \delta g^{\mu\alpha}\nabla_\alpha\chi^\nu + g^{\mu\alpha}\delta\bigl(\nabla_\alpha\chi^\nu\bigr) \notag \\
	&= \delta g^{\mu\alpha}\nabla_\alpha\chi^\nu + \nabla^\mu\delta\chi^\nu + g^{\mu\alpha}\delta\Gamma^\nu_{\alpha\beta}\chi^\beta
\end{align}
leading to
\begin{align}
	\If_\V\delta\bigl(\nabla^\mu\chi^\nu\bigr) &= \bigl(\If_\V\delta g^{\mu\alpha}\bigr)\nabla_\alpha\chi^\nu - \delta g^{\mu\alpha}\nabla_\alpha\bigl(\If_\V\chi^\nu\bigr) \notag \\
	&\quad + \nabla^\mu\bigl(\If_\V\delta\chi^\nu\bigr) + g^{\mu\alpha}\Bigl(\bigl(\If_\V\delta\Gamma^\nu_{\alpha\beta}\bigr)\chi^\beta - \delta\Gamma^\nu_{\alpha\beta}\bigl(\If_\V\chi^\beta\bigr)\Bigr) \notag \\
	&= -\bigl(\nabla^\mu\epsilon^\alpha + \nabla^\alpha\epsilon^\mu\bigr)\nabla_\alpha\chi^\nu + \bigl(\nabla_\alpha\epsilon^\nu\bigr)\delta g^{\mu\alpha} \notag \\
	&\quad + \nabla^\mu\bigl({\cal L}_\epsilon\chi^\nu\bigr) + g^{\mu\alpha}\epsilon^\beta\delta\Gamma^\nu_{\alpha\beta} + g^{\mu\alpha}\bigl(\If_\V\delta\Gamma^\nu_{\alpha\beta}\bigr)\chi^\beta.
\end{align}
where we have noted that
\begin{equation}
	\If_\V\delta\chi^\mu = -\bigl(\If_\V\chi^\nu\bigr)\partial_\nu\chi^\mu + \chi^\nu\partial_\nu\bigl(\If_\V\chi^\mu\bigr) = \epsilon^\nu\partial_\nu\chi^\mu - \chi^\nu\partial_\nu\epsilon^\mu = {\cal L}_\epsilon\chi^\mu.
	\label{Idchi}
\end{equation}
Combining \eqref{deta} and \eqref{Idxi}, we now find (note that each term in $-\delta\xi^{\mu\nu}$ is cancelled against a corresponding term in $\If_\V\delta\eta^{\mu\nu}$)
\begin{align}
	-\delta\xi^{\mu\nu} + \If_\V\delta\eta^{\mu\nu} = \sqrt{-g} \Bigl(&\bigl(\nabla_\alpha\epsilon^\alpha\bigr)\bigl(\nabla^\mu\chi^\nu\bigr) - \bigl(\nabla^\mu\epsilon^\alpha + \nabla^\alpha\epsilon^\mu\bigr)\bigl(\nabla_\alpha\chi^\nu\bigr) \notag \\
	&+ \nabla^\mu\bigl({\cal L}_\epsilon\chi^\nu\bigr) + g^{\mu\alpha}\bigl(\If_\V\delta\Gamma^\nu_{\alpha\beta}\bigr)\chi^\beta\Bigr).
\end{align}
On the second line, we have
\begin{align}
	\nabla^\mu\bigl({\cal L}_\epsilon\chi^\nu\bigr) &= \nabla^\mu\bigl(\epsilon^\alpha\nabla_\alpha\chi^\nu - \chi^\alpha\nabla_\alpha\epsilon^\nu\bigr) \notag \\
	&= \bigl(\nabla^\mu\epsilon^\alpha\bigr)\bigl(\nabla_\alpha\chi^\nu\bigr) + \epsilon^\alpha\nabla^\mu\nabla_\alpha\chi^\nu - \bigl(\nabla^\mu\chi^\alpha\bigr)\bigl(\nabla_\alpha\epsilon^\nu\bigr) - \chi^\alpha\nabla^\mu\nabla_\alpha\epsilon^\nu
\end{align}
and the contraction
\begin{equation}
	\If_\V\delta\Gamma^\mu_{\alpha\beta} = \frac{1}{2}\Bigl(\bigl[\nabla_\alpha, \nabla^\mu\bigr]\epsilon_\beta + \bigl[\nabla_\beta, \nabla^\mu\bigr]\epsilon_\alpha + \bigl(\nabla_\alpha\nabla_\beta + \nabla_\beta\nabla_\alpha\bigr)\epsilon^\mu\Bigr)
\end{equation}
which has already been encountered in our earlier calculations. Hence
\begin{align}
	-\delta\xi^{\mu\nu} + \If_\V\delta\eta^{\mu\nu} = \sqrt{-g} \Bigl(&\bigl(\nabla_\alpha\epsilon^\alpha\bigr)\bigl(\nabla^\mu\chi^\nu\bigr) - \bigl(\nabla^\mu\epsilon^\alpha + \nabla^\alpha\epsilon^\mu\bigr)\bigl(\nabla_\alpha\chi^\nu\bigr) \notag \\
		&+ \bigl(\nabla^\mu\epsilon^\alpha\bigr)\bigl(\nabla_\alpha\chi^\nu\bigr) - \bigl(\nabla^\mu\chi^\alpha\bigr)\bigl(\nabla_\alpha\epsilon^\nu\bigr) - \bigl(\nabla^\mu\nabla_\alpha\epsilon^\nu\bigr)\chi^\alpha + \epsilon^\alpha\bigl(\nabla^\mu\nabla_\alpha\chi^\nu\bigr) \notag \\
		&+ \frac{1}{2}g^{\mu\alpha}\bigl(\nabla_\alpha\nabla^\nu\epsilon_\beta - \nabla^\nu\nabla_\alpha\epsilon_\beta + \nabla_\beta\nabla^\nu\epsilon_\alpha - \nabla^\nu\nabla_\beta\epsilon_\alpha\bigr)\chi^\beta \notag \\
	&+ \frac{1}{2}g^{\mu\alpha}\bigl(\nabla_\alpha\nabla_\beta\epsilon^\nu + \nabla_\beta\nabla_\alpha\epsilon^\nu\bigr)\chi^\beta\Bigr).
	\label{first_two}
\end{align}
The terms on the last two lines can be arranged as
\begin{align}
	&\frac{1}{2}\bigl(\nabla^\mu\nabla^\nu\epsilon_\alpha - \nabla^\nu\nabla^\mu\epsilon_\alpha + \nabla_\alpha\nabla^\nu\epsilon^\mu - \nabla^\nu\nabla_\alpha\epsilon^\mu\bigr)\chi^\alpha + \frac{1}{2}\bigl(\nabla^\mu\nabla_\alpha\epsilon^\nu + \nabla_\alpha\nabla^\mu\epsilon^\nu\bigr)\chi^\alpha \notag \\
	&= \Bigl(\nabla^{[\mu}\nabla^{\nu]}\epsilon_\alpha + \nabla^{[\mu}\nabla_\alpha\epsilon^{\nu]} + \nabla_\alpha\nabla^{(\mu}\epsilon^{\nu)}\Bigr)\chi^\alpha.
\end{align}
Since this will be contracted with the antisymmetric $\dd\sigma_{\mu\nu}$, the last term does not contribute and the second term cancels against a term on the second line of \eqref{first_two}. We are then left with
\begin{align}
	-\delta\xi^{\mu\nu} + \If_\V\delta\eta^{\mu\nu} = \sqrt{-g} \Bigl(&\bigl(\nabla_\alpha\epsilon^\alpha\bigr)\bigl(\nabla^\mu\chi^\nu\bigr) - \bigl(\nabla^\alpha\epsilon^\mu\bigr)\bigl(\nabla_\alpha\chi^\nu\bigr) - \bigl(\nabla_\alpha\epsilon^\nu\bigr)\bigl(\nabla^\mu\chi^\alpha\bigr) \notag \\
	&+ \epsilon^\alpha\bigl(\nabla^\mu\nabla_\alpha\chi^\nu\bigr) + \bigl(\nabla^\mu\nabla^\nu\epsilon_\alpha\bigr)\chi^\alpha\Bigr).
	\label{first_two_simplified}
\end{align}
Recall that these were the first two terms of the integrand in \eqref{LQ'}. The last term is
\begin{equation}
	-{\cal L}_\epsilon\eta^{\mu\nu} = -\epsilon^\alpha\nabla_\alpha\eta^{\mu\nu} + \bigl(\nabla_\alpha\epsilon^\mu\bigr)\eta^{\alpha\nu} + \bigl(\nabla_\alpha\epsilon^\nu\bigr)\eta^{\mu\alpha} - \bigl(\nabla_\alpha\epsilon^\alpha\bigr)\eta^{\mu\nu}.
	\label{last_term}
\end{equation}
Here the last term on the right-hand side is cancelled by the first term in \eqref{first_two_simplified}. The two middle terms can be written as
\begin{align}
	&\bigl(\nabla_\alpha\epsilon^\mu\bigr)\eta^{\alpha\nu} + \bigl(\nabla_\alpha\epsilon^\nu\bigr)\eta^{\mu\alpha} \notag \\
	&= \frac{1}{2}\sqrt{-g}\Bigl(\bigl(\nabla_\alpha\epsilon^\mu\bigr)\bigl(\nabla^\alpha\chi^\nu - \nabla^\nu\chi^\alpha\bigr) + \bigl(\nabla_\alpha\epsilon^\nu\bigr)\bigl(\nabla^\mu\chi^\alpha - \nabla^\alpha\chi^\mu\bigr)\Bigr) \notag \\
	&= \sqrt{-g}\Bigl(\bigl(\nabla^\alpha\epsilon^{[\mu}\bigr)\bigl(\nabla_\alpha\chi^{\nu]}\bigr) + \bigl(\nabla_\alpha\epsilon^{[\nu}\bigr)\bigl(\nabla^{\mu]}\chi^\alpha\bigr)\Bigr)
\end{align}
from which we see that they cancel against the next two terms in \eqref{first_two_simplified} (after taking into account the antisymmetrization due to $\dd\sigma_{\mu\nu}$). Combining now the term remaining in \eqref{last_term}, namely
\begin{equation}
	-\epsilon^\alpha\nabla_\alpha\eta^{\mu\nu} = -\sqrt{-g}\epsilon^\alpha\nabla_\alpha\nabla^{[\mu}\chi^{\nu]}
\end{equation}
with the two remaining terms in \eqref{first_two}, we obtain
\begin{align}
	\Lf_\V\Theta_\phi &= \int_{\partial\Sigma} \dd\sigma_{\mu\nu}\bigl(\phi(\X)\bigr)\,\sqrt{-g} \Bigl(\epsilon^\alpha\bigl(\nabla^\mu\nabla_\alpha\chi^\nu\bigr) + \bigl(\nabla^\mu\nabla^\nu\epsilon_\alpha\bigr)\chi^\alpha - \epsilon^\alpha\bigl(\nabla_\alpha\nabla^\mu\chi^\nu\bigr)\Bigr)\bigl(\phi(\X)\bigr) \notag \\
	&= \int_{\partial\Sigma} \dd\sigma_{\mu\nu}\bigl(\phi(\X)\bigr)\,\sqrt{-g} \Bigl(\epsilon^\alpha\bigl[\nabla^\mu, \nabla_\alpha\bigr]\chi^\nu + \frac{1}{2}\bigl(\bigl[\nabla^\mu, \nabla^\nu\bigr]\epsilon_\alpha\bigr) \chi^\alpha\Bigr)\bigl(\phi(\X)\bigr).
\end{align}
We continue the calculation by expressing the commutators of covariant derivatives in terms of the Riemann tensor. Taking $\mu$ and $\nu$ as lower indices for convenience, we have
\begin{align}
	\bigl[\nabla_\mu, \nabla_\alpha\bigr]\chi_\nu &= R_{\nu\lambda\mu\alpha}\chi^\lambda, \notag \\[1ex]
	\bigl[\nabla_\mu, \nabla_\nu\bigr]\epsilon_\alpha &= R_{\alpha\lambda\mu\nu}\epsilon^\lambda.
\end{align}
and
\begin{equation}
	\epsilon^\alpha\bigl[\nabla_\mu, \nabla_\alpha\bigr]\chi_\nu + \frac{1}{2}\bigl(\bigl[\nabla_\mu, \nabla_\nu\bigr]\epsilon_\alpha\bigr)\chi^\alpha = \epsilon^\alpha R_{\nu\lambda\mu\alpha}\chi^\lambda + \frac{1}{2}R_{\alpha\lambda\mu\nu}\epsilon^\lambda\chi^\alpha  = \Bigl(R_{\nu\lambda\mu\alpha} + \frac{1}{2}R_{\lambda\alpha\mu\nu}\Bigr)\epsilon^\alpha\chi^\lambda.
\end{equation}
Since this expression will be antisymmetrized in $\mu$ and $\nu$, we can write it as
\begin{equation}
	\frac{1}{2}\Bigl(R_{\nu\lambda\mu\alpha} - R_{\mu\lambda\nu\alpha} + R_{\lambda\alpha\mu\nu}\Bigr)\epsilon^\alpha\chi^\lambda.
\end{equation}
Now using the symmetries of the Riemann tensor, namely $R_{\alpha\beta\mu\nu} = -R_{\alpha\beta\nu\mu}$ and $R_{\alpha\beta\mu\nu} = R_{\mu\nu\alpha\beta}$, we see that
\begin{align}
	&R_{\nu\lambda\mu\alpha} - R_{\mu\lambda\nu\alpha} + R_{\lambda\alpha\mu\nu} \notag \\
	&= R_{\mu\alpha\nu\lambda} - R_{\mu\lambda\nu\alpha} + R_{\mu\nu\lambda\alpha} \notag \\
	&= R_{\mu\alpha\nu\lambda} + R_{\mu\lambda\alpha\nu} + R_{\mu\nu\lambda\alpha} \notag \\
	&= 0
\end{align}
due to the first Bianchi identity. Hence we have managed to show that
\begin{equation}
	\Lf_\V\Theta_\phi = 0.
\end{equation}
Note that this result followed from the algebraic properties of the Riemann tensor and no use of the equations of motion was necessary in order to establish it, reflecting the fact that the invariance of the boundary presymplectic form under diffeomorphisms is a purely geometric property, and as such it should be valid independently of any dynamics.

\clearpage

\printbibliography

\end{document}